%% file: Paper1_Jan9.tex
\documentclass[11pt,twoside,a4paper]{article}
\makeatletter\if@twocolumn\PassOptionsToPackage{switch}{lineno}\else\fi\makeatother

      \makeatletter
\usepackage{wrapfig}
\newcounter{aubio}

\long\def\bioItem{%
\@ifnextchar[{\@bioItem}{\@@bioItem}}

\long\def\@bioItem[#1]#2#3{
 \stepcounter{aubio}
 \expandafter\gdef\csname authorImage\theaubio\endcsname{#1}
 \expandafter\gdef\csname authorName\theaubio\endcsname{#2}
 \expandafter\gdef\csname authorDetails\theaubio\endcsname{#3}
}

\long\def\@@bioItem#1#2{
 \stepcounter{aubio}
 \expandafter\gdef\csname authorName\theaubio\endcsname{#1}
 \expandafter\gdef\csname authorDetails\theaubio\endcsname{#2}
}

\newcommand{\checkheight}[1]{%
  \par \penalty-100\begingroup%
  \setbox8=\hbox{#1}%
  \setlength{\dimen@}{\ht8}%
  \dimen@ii\pagegoal \advance\dimen@ii-\pagetotal
  \ifdim \dimen@>\dimen@ii
    \break
  \fi\endgroup}

\def\printBio{%
  \@tempcnta=0
   \loop
     \advance \@tempcnta by 1
     \def\aubioCnt{\the\@tempcnta}
     \setlength{\intextsep}{0pt}%
     \setlength{\columnsep}{10pt}%
     \expandafter\ifx\csname authorImage\aubioCnt\endcsname\relax%
      \else%
       \checkheight{\includegraphics[height=1.25in,width=1in,keepaspectratio]{\csname authorImage\aubioCnt\endcsname}}
        \begin{wrapfigure}{l}{25mm}
         \includegraphics[height=1.25in,width=1in,keepaspectratio]{\csname authorImage\aubioCnt\endcsname}%height=145pt
        \end{wrapfigure}\par
      \fi
     \noindent\textbf{\csname authorName\aubioCnt\endcsname}\csname authorDetails\aubioCnt\endcsname \par\bigskip
      \ifnum\@tempcnta < \theaubio
   \repeat
   }
\makeatother

\usepackage{amsfonts,amssymb,amsbsy,latexsym,amsmath,tabulary,graphicx,times,xcolor}

\usepackage[T1]{fontenc}

\usepackage{caption,fancyhdr,abstract,textcase}
\usepackage[paperheight=297mm,paperwidth=210mm,margin=30mm,left=30mm,right=30mm,top=30mm]{geometry}
\usepackage{textcase}
\makeatletter
\def\oupIndent{1pt}
\def\author#1{\gdef\@author{\hskip-\dimexpr(\tabcolsep)\hskip1pt\parbox{\dimexpr\textwidth-1pt}{\centering{\fontsize{13pt}{15.6pt}\selectfont  #1}}}}

\def\title#1{\gdef\@title{\vspace*{-3pc}\bfseries\centering\ifx\@articleType\@empty\else\@articleType\\\fi {\fontfamily{ppl}\fontsize{20pt}{24pt}\selectfont\MakeTextUppercase{ #1} \vspace*{24pt}}}}

\fancypagestyle{headings}{\fancyhf{}\fancyfoot[R]{}}

\fancypagestyle{plain}{\fancyhf{}\fancyfoot[R]{}}

\let\@articleType\@empty \def\articletype#1{\gdef\@articleType{{\normalfont\underline{#1}}}}

\def\abstractname{\textbf{\textit{A\small{BSTRACT}}}}

\renewenvironment{abstract} {\trivlist\item[]\leftskip\oupIndent\par\vskip4pt\noindent{\fontsize{13pt}{15.6pt}\selectfont\textit{\scshape\abstractname}}\mbox{\null}\vspace{5pt}\\ \itshape\fontsize{10pt}{12pt}\selectfont}{\noindent\endtrivlist}

\usepackage[explicit]{titlesec}
\def\NormalBaseline{\def\baselinestretch{1.1}}
\titleformat{\section}[hang]{\NormalBaseline\filright\large\scshape\bfseries\fontsize{14}{16.8}\selectfont}
{\fontsize{14}{16.8}\selectfont\thesection.}
{2pt}
{#1}
[]
\titleformat{\subsection}[hang]{\NormalBaseline\filright\bfseries\fontsize{12}{14.4}\selectfont}
{\thesubsection.}
{2pt}
{#1}
[]
\titleformat{\subsubsection}[hang]{\NormalBaseline\filright\bfseries\fontsize{11}{13.2}\selectfont}
{\thesubsubsection.}
{2pt}
{#1}
[]
\titleformat{\paragraph}[runin]{\NormalBaseline\filright\itshape\fontsize{11}{13.2}\selectfont}
{\theparagraph}
{2pt}
{#1}
[\unskip.]
\titleformat{\subparagraph}[runin]{\NormalBaseline\filright\fontsize{11}{13.2}\selectfont}
{\thesubparagraph}
{2pt}
{#1}
[\unskip.]

\titlespacing{\section}{0pt}{1.5\baselineskip}{.2\baselineskip}
\titlespacing{\subsection}{0pt}{1.5\baselineskip}{.2\baselineskip}
\titlespacing{\subsubsection}{0pt}{1.5\baselineskip}{.2\baselineskip}
\titlespacing{\paragraph}{0pt}{.5\baselineskip}{10pt}
\titlespacing{\subparagraph}{0pt}{.5\baselineskip}{10pt}

\makeatother

\date{}

\usepackage{url,multirow,morefloats,floatflt,cancel,tfrupee}
\makeatletter

\AtBeginDocument{\@ifpackageloaded{textcomp}{}{\usepackage{textcomp}}}
\makeatother
\usepackage{colortbl}
\usepackage{xcolor}
\usepackage{pifont}
\usepackage[nointegrals]{wasysym}
\makeatletter

\def\mcWidth#1{\csname TY@F#1\endcsname+\tabcolsep}

\def\cAlignHack{\rightskip\@flushglue\leftskip\@flushglue\parindent\z@\parfillskip\z@skip}
\def\rAlignHack{\rightskip\z@skip\leftskip\@flushglue \parindent\z@\parfillskip\z@skip}

\usepackage{ifxetex}
\ifxetex\else\if@twocolumn\usepackage{dblfloatfix}\fi\fi

\AtBeginDocument{
\expandafter\ifx\csname eqalign\endcsname\relax
\def\eqalign#1{\null\vcenter{\def\\{\cr}\openup\jot\m@th
  \ialign{\strut$\displaystyle{##}$\hfil&$\displaystyle{{}##}$\hfil
      \crcr#1\crcr}}\,}
\fi
}

\AtBeginDocument{%
  \@ifpackageloaded{endfloat}%
   {\renewcommand\efloat@iwrite[1]{\immediate\expandafter\protected@write\csname efloat@post#1\endcsname{}}}{\newif\ifefloat@tables}%
}%

\def\BreakURLText#1{\@tfor\brk@tempa:=#1\do{\brk@tempa\hskip0pt}}
\let\lt=<
\let\gt=>
\def\processVert{\ifmmode|\else\textbar\fi}

\@ifundefined{subparagraph}{
\def\subparagraph{\@startsection{paragraph}{5}{2\parindent}{0ex plus 0.1ex minus 0.1ex}%
{0ex}{\normalfont\small\itshape}}%
}{}

\newcommand\role[1]{\unskip}
\newcommand\aucollab[1]{\unskip}

\@ifundefined{tsGraphicsScaleX}{\gdef\tsGraphicsScaleX{1}}{}
\@ifundefined{tsGraphicsScaleY}{\gdef\tsGraphicsScaleY{.9}}{}
\def\checkGraphicsWidth{\ifdim\Gin@nat@width>\linewidth
	\tsGraphicsScaleX\linewidth\else\Gin@nat@width\fi}

\def\checkGraphicsHeight{\ifdim\Gin@nat@height>.9\textheight
	\tsGraphicsScaleY\textheight\else\Gin@nat@height\fi}

\def\fixFloatSize#1{}%\@ifundefined{processdelayedfloats}{\setbox0=\hbox{\includegraphics{#1}}\ifnum\wd0<\columnwidth\relax\renewenvironment{figure*}{\begin{figure}}{\end{figure}}\fi}{}}
\let\ts@includegraphics\includegraphics

\def\inlinegraphic[#1]#2{{\edef\@tempa{#1}\edef\baseline@shift{\ifx\@tempa\@empty0\else#1\fi}\edef\tempZ{\the\numexpr(\numexpr(\baseline@shift*\f@size/100))}\protect\raisebox{\tempZ pt}{\ts@includegraphics{#2}}}}

\AtBeginDocument{\def\includegraphics{\@ifnextchar[{\ts@includegraphics}{\ts@includegraphics[width=\checkGraphicsWidth,height=\checkGraphicsHeight,keepaspectratio]}}}

\DeclareMathAlphabet{\mathpzc}{OT1}{pzc}{m}{it}

\def\URL#1#2{\@ifundefined{href}{#2}{\href{#1}{#2}}}

\def\UrlOrds{\do\*\do\-\do\~\do\'\do\"\do\-}%
\g@addto@macro{\UrlBreaks}{\UrlOrds}

\@ifundefined{quoteAttrib}
	{}
	{}

\@ifundefined{titlequoteAttrib}
	{}{}

\newenvironment{title-quote}
	{\list{}{\fontsize{10pt}{12pt}\selectfont\leftmargin.5in\itshape\rightmargin\leftmargin}%
  \item\relax}
  {\endlist}

\makeatother

\usepackage[numbers,sort&compress]{natbib}

\usepackage{supertabular}
\usepackage{graphicx}
\usepackage{longtable}
\usepackage{amsmath}
\usepackage{tabularx}

\usepackage{caption}
\usepackage{subcaption}

\usepackage{chronology}
\usepackage{algorithmic}
\usepackage[]{algorithm2e}
\usepackage{amssymb}
\usepackage{float}
\usepackage{threeparttable}

\usepackage{array}
\usepackage{subcaption}
\usepackage{multirow}
\usepackage{rotating}
\usepackage{pdflscape}

\usepackage{amssymb}

\begin{document}

%\title{\mbox{}}
\title{\huge{Service Level Driven Job Scheduling in Multi-Tier Cloud Computing: A Biologically Inspired Approach}}
\author{\Large{Husam Suleiman~and~Otman Basir}\\
\large{Department of Electrical and Computer Engineering, University of Waterloo}\\
\large{hsuleima@uwaterloo.ca,~obasir@uwaterloo.ca}}

\def\journalTitle{International Journal on Cloud Computing: Services and Architecture (IJCCSA)}

\maketitle

\begin{abstract}
Cloud computing environments often have to deal with random-arrival computational workloads that vary in resource requirements and demand high Quality of Service (QoS) obligations. It is typical that a Service-Level-Agreement (SLA) is employed to govern the QoS obligations of the cloud computing service provider to the client. A typical challenge service-providers face every day is maintaining a balance between the limited resources available for computing and the high QoS requirements of varying random demands. Any imbalance in managing these conflicting objectives may result in either dissatisfied clients and potentially significant commercial penalties, or an over-resourced cloud computing environment that can be significantly costly to acquire and operate. Thus, scheduling the clients' workloads as they arrive at the environment to ensure their timely execution has been a central issue in cloud computing. Various approaches have been reported in the literature to address this problem: Shortest-Queue, Join-Idle-Queue, Round Robin, MinMin, MaxMin, and Least Connection, to name a few. However, optimization strategies of such approaches fail to capture QoS obligations and their associated commercial penalties. %Solution:
This paper presents an approach for service-level driven load scheduling and balancing in multi-tier environments. Joint scheduling and balancing operations are employed to distribute and schedule jobs among the resources, such that the total waiting time of client jobs is minimized, and thus the potential of a penalty to be incurred by the service provider is mitigated. A penalty model is used to quantify the penalty the service provider incurs as a function of the jobs' total waiting time. A Virtual-Queue abstraction is proposed to facilitate optimal job scheduling at the tier level. This problem is NP-complete, thus, a genetic algorithm is proposed as a tool for computing job schedules that minimize potential QoS penalties and hence minimize the likelihood of dissatisfied clients.
%
%This paper presents an approach for service-level driven load scheduling and balancing in multi-tier environments. A penalty model is used to quantify the amount of penalty the service provider incurs as a function of the jobs' total waiting time. Joint scheduling and balancing operations are employed to distribute and schedule jobs among the resources, such that the total waiting time and thus QoS penalty of client jobs are minimized. A Virtual-Queue abstraction is proposed to facilitate optimal job scheduling at the tier level. This problem is NP-complete, thus a genetic algorithm is proposed as a tool that the service provider can use to compute scheduling and load balancing decisions that minimize the likelihood of dissatisfied clients.
\end{abstract}
\emph{\textbf{KEYWORDS}}\\
\emph{\small{Heuristic Optimization, Job Scheduling, Job Allocation, Load Balancing, Multi-Tier Cloud Computing}}
%The paper reports experimental results to demonstrate the efficacy of the proposed technique. It is shown that the proposed approach is more effective in minimizing the total waiting time (or SLA penalties) of client jobs (incurred by the service provider) compared to existing techniques. % also minimize penalty!!!

\input{Introduction_Jan9}

\input{Literature_Jan9}

\input{Problem_Jan9}

\input{GeneticAlgorithm_Jan9}

\input{Results_Jan9}

\input{Conclusion_Jan9}

\bibliographystyle{IEEEtran}

\bibliography{\jobname} % this is enough, however you just need to name your bib file similar to this doc's name
%\bibliography{Paper1_Jan9}
\end{document}

%% file: Introduction_Jan9.tex
\section{Introduction}
\label{sec:intro}

The cloud computing is gaining momentum as the computing environment of choice that leverages a set of existing technologies to deliver better service and meet varying demands of clients. Cloud services are provided to clients as software-as-a-service, platforms, and infrastructures. Such services are accessed over the Internet using broad network connections. The success of cloud computing, in all cases, hinges largely on the proper management of the cloud resources to achieve cost-efficient job execution and high client satisfaction. The virtualization of computing resources is an important concept for achieving this goal~\cite{1_cloudRaouf_2010, cloudSurv2012}.

Generally speaking, cloud computing jobs vary in computational needs and QoS requirements. They can be can be periodic, aperiodic, or sporadic~\cite{JobSizeSched_2000}. Each job has a prescribed execution time and tardiness allowance. Typically, an SLA is employed to govern the QoS expectations of the client, as well as a model to compute penalties in cases of QoS violation~\cite{SoftHardSLA, SLAlevels}. A major challenge cloud computing service providers face is maintaining a maximum resource utilization while ensuring adequate resource availability to meet the QoS expectation in executing jobs of varying computational demands. Failing to meet its clients QoS demands may result into harsh financial penalties and dissatisfaction of customers. On the other hand, procuring large assets of computational resources can be prohibitively costly. Thus, it is imperative that jobs are allocated to resources and scheduled for processing so as to minimize their waiting time in the environment. The goodness of any scheduling strategy hinges on its ability to achieve high computing performance and maximum system throughput.

There are two job scheduling categories, namely, preemptive scheduling and non-preemptive scheduling. Under preemptive scheduling, a running job can be taken away from the server or interrupted to accommodate a higher priority job. Furthermore, non-preemptive scheduling, in general, is simpler to implement, no synchronization overhead, minimum stack memory needs. Under non-preemptive scheduling, a running job cannot be taken away from the CPU or interrupted. While non-preemptive scheduling allows for a predictable system response time, preemptive scheduling, on the other hand, emphasizes real-time response time at the job level.

In this paper, we are concerned with the issue of job scheduling in multi-tier cloud computing environments.  Given a set of client jobs with  different computational demands/constraints, to be executed on a multi-tier cloud computing environment, it is desirable that these jobs are scheduled for execution in this environment such that the penalty to be incurred by the service provider is minimized. It is assumed that the environment consists of a set of cascaded tiers of identical computing resources. 

Emphasizing the notion of penalty in scheduling the jobs allows us to imply job priority of treatment that is based on economic considerations. As such, the service provider is able to leverage available job tardiness allowances and QoS penalty considerations to compute schedules that yield minimum total penalty. This strategy is particularly useful in situations of excessive volume of demands or lack of an adequate resource availability that will make it impossible to meet the QoS requirements.

%The paper organization is as follows. Section~\ref{sec:backRelated} presents the background and related work. Section~\ref{sec:probFormal} explains the problem formalization. Sections~\ref{sec:QueueVirt} and \ref{sec:penaltyAppr} clarify the queue virtualization and penalty-driven tier-based genetic solution used to tackle the problem, respectively. The results are discussed in Section~\ref{sec:discResults}, followed by the conclusion and future directions in Sections~\ref{sec:conc} and \ref{sec:future}, respectively.

%% file: Literature_Jan9.tex
\section{Background and Related Work}
\label{sec:backRelated}

The issue of job scheduling has been an active area of research since the early days of computing.
A large body of work exists in the literature about scheduling approaches in cloud computing environments \cite{A_Comparative_Survey_2014, LBghomiSurv2017, TaskSurv_2019}.
Such approaches, generally speaking, aim at minimizing the average response time of jobs and maximizing resource utilization. Schroeder \emph{et al}.~\cite{Bianca_2004} evaluate the Least-Work-Left, Random, and Size-Interval based Task Assignment (SITA-E) scheduling approaches on a single-tier environment that consists of distributed resources with one dispatcher. They propose an approach that purposely unbalances the load between resources. The mean response time and slowdown metrics are used to assess each approach. The primary drawback in their work is that they don't consider migrating jobs between resource queues and as a result fail to produce optimal schedules.

Liu \emph{et al}.~\cite{MinMin_2013} report a Min-Min algorithm for task scheduling that makes jobs with long times execute at reasonable times. Li \emph{et al}.~\cite{MaxMin_2014} present a Max-Min scheduling approach that estimates the total workload and completion time of jobs in each resource, so as to allocate jobs on resources to reduce their average response time. In both approaches, the scheduling decisions dedicate powerful resources to execute specific jobs without accurately considering the different QoS expectations of jobs. For instance, a Max-Min approach assigns the job with the maximum execution time to the resource that provides the minimum completion time for the job, yet does not account for the different job constraints and the impacts of their violations on the QoS. However, states of resource queues are not considered when the decisions are taken, and accordingly ineffective distribution of workloads among the resource queues is expected to occur. Furthermore, such approaches tackle jobs that mainly arrive in batches. When jobs of different constraints and requirements arrive in a consecutive/dynamic manner to a multi-tier cloud computing environment, the scheduling decisions of such approaches would fail to accurately capture the QoS obligations and economical impacts of these jobs on the service provider and client.

Some approaches take advantage of the knowledge obtained about the system state to make scheduling decisions. Examples of these approaches are: Least Connect (LC), Join-Shortest-Queue (JSQ), Weighted Round Robin (WRR), and Join-Idle-Queue (JIQ). Gupta \emph{et al}.~\cite{Mor_2011} present and analyze the JSQ approach in a farm of servers, that is similar in architecture to a single-tier cloud environment. JSQ assumes the resource of the least number of jobs is the least loaded resource. In contrast, the weighted algorithms (e.g., WRR and WLC) are commonly used in balancing the load among resources in cloud environments~\cite{WLC_2015,WRR_2014}. Wang \emph{et al}.~\cite{WRR_2014} effectively apply the WRR algorithm, by determining weights for resources based on their computational capabilities, then allocating and balancing the workloads among these resources. Though, powerful resources would receive extra workloads of jobs. However, the states of the resource queues are not accurately measured, and thus scheduling decisions taken based on only weights of resources often lead to load unbalance among the resource queues.

Lu \emph{et al}.~\cite{JIQ_2011} present a JIQ algorithm for large-scale load balancing problems to minimize the communication overhead incurred between resources and multiple distributed dispatchers at the time of job arrivals. In the JIQ algorithm, each dispatcher has a separate idle queue to maintain IDs of idle resources of the tier. An idle resource informs specific dispatcher(s) of its availability to receive jobs. The primary drawback is that an idle resource might experience significant queuing bottlenecks when it requests jobs from many dispatchers at the same time. Also, an idle resource might run the risk of not receiving jobs and thus get under-utilized if its associated/informed dispatchers are empty of jobs, while at the same time other uninformed dispatchers of the tier might be holding jobs waiting to get idle resources. Sometimes, low priority jobs might get a dispatcher that has IDs of idle resources, whilst high priority jobs might be assigned to a dispatcher that has not yet held signals of idle resources. In this case, the low priority jobs would get scheduled and executed in idle resources while the high priority jobs are still waiting.

The Round Robin algorithm, which has been popular in process scheduling, has been adopted in cloud computing to tackle the job scheduling problem. The Round Robin algorithm aims at distributing the load equally to all resources~\cite{Pooja}. Using this algorithm, one Virtual Machine (VM) is allocated to a node in a cyclic manner. However, the Round Robin algorithms are based on a simple cyclic scheduling scenario, thus lead to unbalancing the traffic and incur more load on resources. In general, Round Robin algorithms have shown improved response times and load balancing.

Some researchers have adopted Bio-inspired meta-heuristic approaches to tackle the scheduling problem in cloud environments~\cite{heuristic1, ACO_MaxMin_2015, PSO_2014}. For instance, Nouiri \emph{et al}.~\cite{heuristicPSO} present a particle swarm optimization algorithm to minimize the maximum makespan. Nevertheless, their bio-inspired job scheduling formulation makes scheduling decisions to benefit specific jobs at the expense of other jobs. Such approaches disregard economical penalties that may result from scheduling decisions. Instead, they focus on optimizing system-level metrics. Job response time, resource utilization, maximum tardiness, and completion time are typically used metrics.

Mateos \emph{et al}.~\cite{Mateos2013AnAA} propose an Ant Colony Optimization (ACO) approach to implement a scheduler for cloud computing applications, which aims at minimizing the weighted flow-time and Makespan of jobs. The load is calculated on each resource, taking into consideration CPU utilization of all the VMs that are executing on each host. CPU utilization is used as a metric that allows Ant to choose the least loaded host to allocate its VM. Pandey \emph{et al}.~\cite{Pandey2010APS} report a Particle Swarm Optimization (PSO) algorithm for minimizing the computational cost of application workflow in cloud computing environments. Job execution time is used as a performance metric. The PSO based resource mapping demonstrated superior performance when compared to Best Resource Selection (BRS) based mapping. Furthermore, the PSO based algorithm achieved optimal load balance among the computing resources.

As a general observation, current cloud computing approaches contemplate single tier environments and fail to exploit resource queue dynamics to migrate jobs between the resources of a given tier so as to achieve the optimal job scheduling. Furthermore, schedule optimality is defined based on job response time metrics. The reality is, the scheduling problem is an NP problem and there are situations where finding schedules that satisfy the target response times of all jobs is an impossible task due to resource limitations, even if we are to put the problem complexity aside. Cloud computing clients are not the same with respect to their QoS expectations.

Furthermore, the impact of the job execution violation on the QoS differs from job to another. There are computing jobs that can tolerate some degree of execution violation, however, there are other jobs that are tightly coupled with mission-critical obligations or user experience. These jobs tend to be intolerant to execution delays. SLAs tend to provide a context based on which differential job treatment regimes can be devised. The impact of job violation on QoS tends to be captured in a penalty model. In this work we propose to leverage this model to influence scheduling in a multi-tier cloud computing environment so as to minimize penalty payable by the service provider and as a result attain a pragmatic QoS.

%% file: Problem_Jan9.tex
\begin{figure}[!t]
\centering
%\begin{flushleft}
	  \includegraphics[width=0.75\textwidth,height=0.3375\textheight]{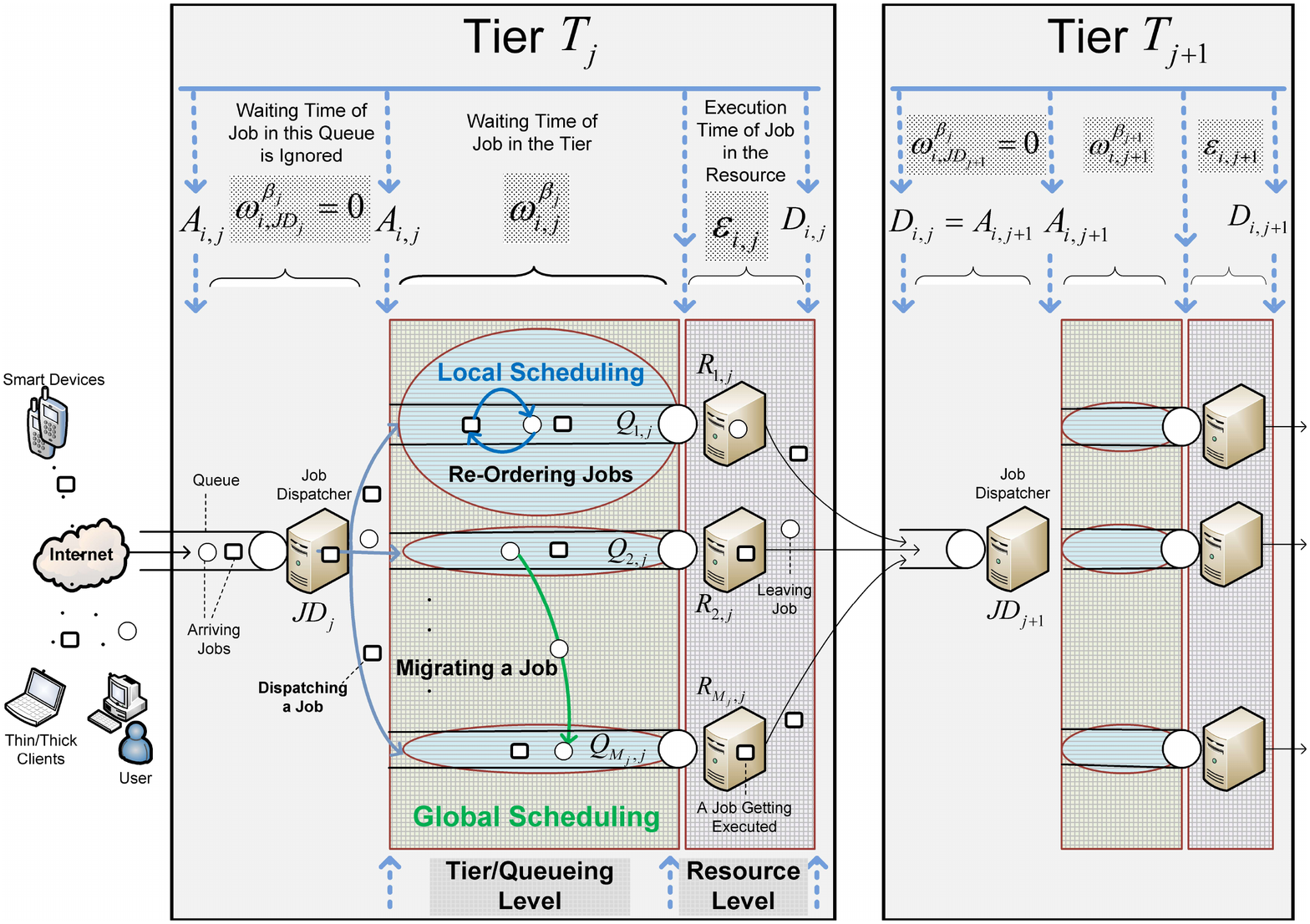}
	  \caption{System Model of the Multi-Tier Environment}
      \label{fig:Paper1_SystemModel1_1}
%\end{flushleft}
\end{figure} %(default)width=\textwidth,height=0.45\textheight of whole page
% width=0.475\textwidth,height=0.2565\textheight

\section{SLA Driven Job Scheduling}
\label{sec:probFormal}

We consider a multi-tier cloud computing environment consisting of $N$ sequential tiers:
\begin{equation}
T = \left \{T\!_1,T\!_2,T\!_3,...,T\!_N  \right \}
\end{equation}

Each tier $T\!_j$ employs a set of identical computing resources $R_j$:
\begin{equation}
R_j = \left \{R_{1,j},R_{2,j},R_{3,j},...,R_{M_j,j}  \right \}
\end{equation}

Each resource $R_{k,j}$ employs a queue $Q_{k,j}$ to hold jobs waiting for execution by the resource.
Jobs with different computational requirements are submitted to the environment. It is assumed that these jobs are submitted by different clients and hence are governed by various SLA's. Jobs arrive at the environment in streams. 
% A stream $S$ is a set of jobs:
%\begin{equation}
%S = \left \{ J_1,J_2,J_3,...,J_l  \right \}
%\end{equation}
Job $J_i$ arrives at the environment before job $J_{i+1}$. Jobs arrive in a random manner. $A_{i,j}$ is the arrival time of Job $J_i$ at tier $T_j$; $\mathcal{E}_{i,j}$ is its prescribed execution time at this tier.

%\label{equ:Ji}
%    J_i = \left \{A_{i,j}, \mathcal{E}_{i,j} \right \},\;\;\; \forall\; T_j\!\in\!T
%\end{equation}

Jobs arriving at tier $T_j$ are queued for execution based on an ordering $\beta_j$. As shown in Figures~\ref{fig:Paper1_SystemModel1_1} and \ref{fig:tierVirtualQueue}, each tier $T_j$ of the environment consists of a set of resources. Each resource has a queue to hold jobs assigned to it. For instance, resource $R_{1,j}$ of tier $T_j$ is associated with queue $Q_{1,j}$, which consists of $4$ jobs ($J_6$, $J_7$, $J_8$, and $J_{10}$) waiting for execution. A virtual-queue is a cascade of all queues of a given tier.
The total execution time $E_i$ of  job $J_i$ is defined as
\begin{equation}
\label{equ:Ei}
    E_i = \sum_{j=1}^{N} \mathcal{E}_{i,j}
\end{equation}

Job $J_i$'s response time $Z_i$ is a function of its total execution time $E_i$ and waiting time $W_i$, which in turn is dependent on its position in the queues as it progresses from one tier to the next:
\begin{equation}
\label{equ:Zi}
    Z_i = \sum_{j=1}^{N} (\mathcal{E}_{i,j} + \omega_{i,j}^{\beta_j}) = E_i + W_i
\end{equation}
where $\omega_{i,j}^{\beta_j}$ represents the waiting time of job $J_i$ at tier $T_j$; $\beta_j$ is the ordering that governs the order of execution of jobs at tier $T_j$. $W_i$ is the total time job $J_i$ waits at all tiers. Job $J_i$ departs $T_j$ at time $D_{i,j}$.
% which will be the arrival time $A_{i,j+1}$ of the job to the next tier $T_{j+1}$.
%\begin{equation}
%\label{equ:beta}
%    \beta = \bigcup_{j=1}^{N} \beta_j
%\end{equation}

%The execution time $\mathcal{E}_{i,j}$ of job $J_i$ at tier $T_j$ is pre-defined in advance. Therefore, the resource capabilities of each tier $T_j$ are not considered and thus the total execution time $\mathcal{E\!T}\!\!_i$ of job $J_i$ is constant.

%The primary concern is on the queueing-level of the environment represented by the total waiting time $\omega\!\mathcal{T}_i$ of job $J_i$ at all tiers $T$.% according to the ordering $\beta$.

We assume a service level agreement that stipulates a penalty amount as an exponential function of the difference between the prescribed response time and the actual response time.
%The response time of job $J_i$ in the environment is subject to an SLA that stipulates an exponential penalty curve $\varrho_i$ as follows:
\begin{equation}
\label{equ:rhoi}
\begin{split}
    \varrho_i & = \chi * (1-\text{e}^{-\nu(Z\!_i-E_i)}) \\
              & = \chi * ( 1- \text{e}^{- \nu(W_i )} ) \\
              & = \chi * ( 1- \text{e}^{- \nu\sum_{j=1}^{N} \omega_{i,j}^{\beta_j} } )
\end{split}
\end{equation}
where $\chi$ is a monitory cost factor and $\nu$ is an arbitrary scaling factor. The total penalty cost associated with a set of jobs $J_1\dots J_l$ , across all tiers is defined as:
\begin{equation}
\label{equ:Phi}
    \varphi = \sum_{i=1}^{l} \varrho_i
\end{equation}

%The total penalty cost $\varphi$ includes calculating the waiting time $\omega_{i,j}^{\beta_j}$ of each job $J_i$ according to the ordering $\beta_j$ of all jobs at tier $T_j$ as follows:
%\begin{equation}
%\label{equ:betaj}
%    \beta_j = \bigcup_{k=1}^{M_k} \text{I}(Q_{k,j})
%\end{equation}
%\begin{equation}
%\label{equ:wijbetaj}
%    \omega_{i,j}^{\beta_j} = \sum_{h\in \text{I}(Q_{k,j}),\;h\;\text{precedes job}\;J_i}^{\forall} \mathcal{E}_{h,j}
%\end{equation}

%where $\text{I}(Q_{k,j})$ represents indices of jobs in $Q_{k,j}$. For instance, $\nolinebreak{\text{I}(Q_{1,2})=\{3,5,2,7\}}$ signifies that jobs $J_3$, $J_5$, $J_2$, and $J_7$ are queued in $Q_{1,2}$ such that job $J_3$ precedes job $J_5$ which in turn precedes job $J_2$, and so on.

The objective is to find optimal set of orderings $\nolinebreak{\beta=(\beta_1, \beta_2, \beta_3, \ldots, \beta_N)}$ such that the total penalty cost $\varphi$ is minimum:
\begin{equation}
\label{equ:min}
\begin{split}
     \underset{\beta}{\text{minimize}} (\varphi) & \equiv \underset{\beta}{\text{minimize}} (\sum_{i=1}^{l} \sum_{j=1}^{N} \omega_{i,j}^{\beta_j})
\end{split}
\end{equation}

$\beta_j$ is an ordering of the jobs waiting at the virtual queue of tier $T_j$.

%% file: GeneticAlgorithm_Jan9.tex
\section{Minimum Penalty Job Scheduling: A Genetic Algorithm Formulation}
\label{sec:penaltyAppr}

The paper is concerned with scheduling the client jobs for execution on the computing resources. A job is first submitted to tier-1 for execution by one of the resources of the tier. It is desired that the jobs are scheduled in such a way that minimizes the total waiting time. Finding a job scheduling that yields minimum total waiting time is an NP problem. Given the volume of cloud jobs to be scheduled and the computational complexity of the job scheduling problem, it is prohibitive to seek an optimal solution for the job scheduling problem using exhaustive search techniques. Thus, a meta-heuristic search strategy, such as Permutations Genetic Algorithms (PGA), is a viable option for exploring and exploiting the large space of scheduling permutations~\cite{GA3}. Genetic algorithms have been successfully adopted in various problem domains~\cite{GaTabu1}. They have undisputed success in yielding near optimal solutions for large scale problems, in reasonable time~\cite{heuristicPSO}.
\begin{figure}[!t]
\centering
%\vspace{-5mm}
%\begin{flushleft}
	  \includegraphics[width=0.4\textwidth,height=0.304\textheight]{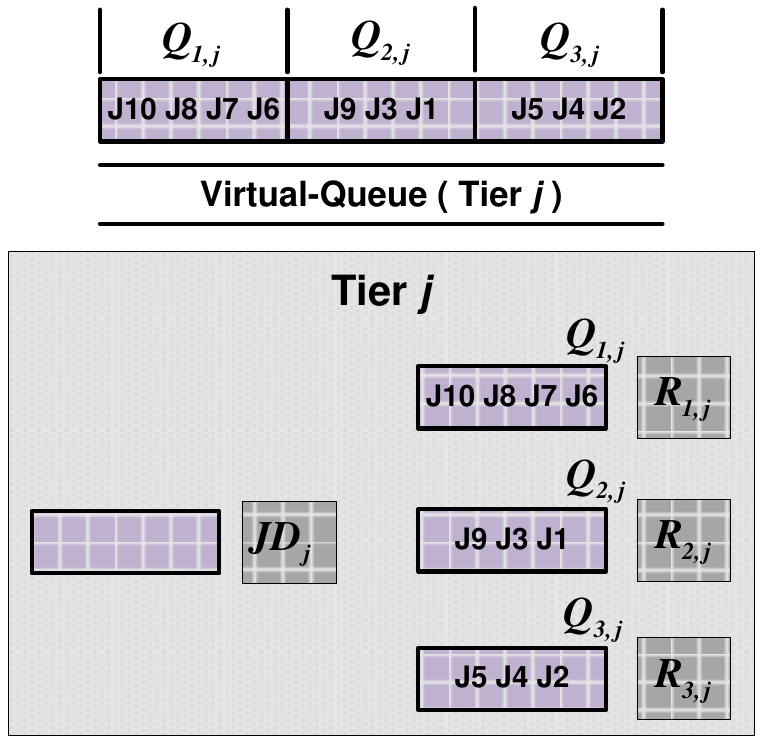}
      %\vspace{-1mm}
	  \caption{The Virtual-Queue of a Tier $j$}
      \label{fig:tierVirtualQueue}
%\end{flushleft}
%\vspace{-3mm}
\end{figure} %(default)width=0.5\textwidth,height=0.38\textheight of whole page and varies when you change the figure
%width=0.33\textwidth,height=0.2508\textheight, it was width=0.275\textwidth,height=0.209\textheight

Scheduling cloud jobs entails two steps: 1) allocating/distributing the jobs among the different tier resources. Jobs that are allocated to a given resource are queued in the queue of that resource; 2) ordering the jobs in the queue of the resource such that their total waiting time is minimum. What makes the problem increasingly harder is the fact that jobs continue to arrive at the tier, while the prior jobs are waiting in their respective queues for execution. Thus, the scheduling process needs to respond to the job arrival dynamics to ensure that job execution at all tiers remains waiting-time optimal. To achieve this, job ordering in each queue should be treated as a continuous process. Furthermore, jobs should be migrated from one queue to another so as to ensure balanced job allocation and maximum resource utilization. Thus, we introduce two operators for constructing optimal job schedules at the tier level:
\begin{itemize}
  \item The \emph{reorder} operator is used to change the ordering of jobs in a given queue so as to find an ordering that minimizes the total waiting time of all jobs in the queue.
  \item The \emph{migrate} operator, on the other hand, is used to exploit the benefits of moving jobs between the different resources of the tier so as to reduce the total waiting time at the tier level. This process is adopted at each tier of the environment.
\end{itemize}

\begin{figure}[!t]
\centering
	  \includegraphics[width=0.8\textwidth,height=0.656\textheight]{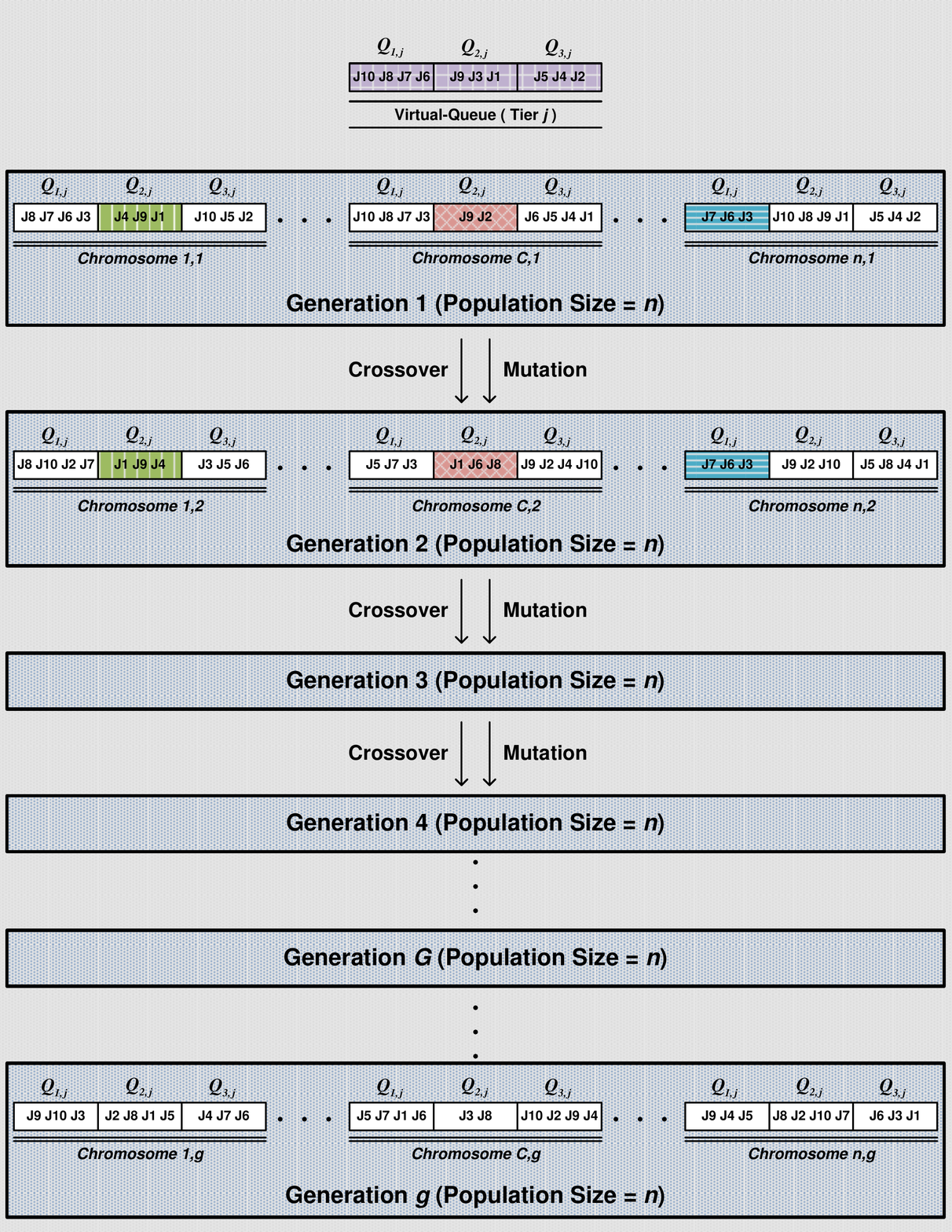}
	  \caption{A Tier-based Genetic Approach on the Virtual-Queue}
      \label{fig:GA}
\end{figure}

However, implementing the \emph{reorder/migrate} operators in a PGA search strategy is not a trivial task. This implementation complexity can be relaxed by virtualizing the queues of each tier into one virtual queue. The virtual queue is simply a cascade of the queues of the resources of the tier. In this way we converge the two operators into simply a reorder operator. Furthermore, this simplifies the PGA solution formulation. A consequence of this abstraction is the length of the permutation chromosome and the associated computational cost. This virtual queue will serve as the chromosome of the solution. An index of a job in this queue represents a gene. The ordering of jobs in a virtual queue signifies the order at which the jobs in this queue are to be executed by the resource associated with that queue. Solution populations are created by permuting the entries of the virtual queue, using the \emph{order} and \emph{migrate} operators.
The virtual-queue in Figures~\ref{fig:tierVirtualQueue} and \ref{fig:GA} of the $j^{\text{th}}$ tier has three queues ($Q_{1,j}$, $Q_{2,j}$, and $Q_{3,j}$) cascaded to construct one virtual queue.

\subsection{Evaluation of Schedules}
\label{sec:step3}

A fitness evaluation function is used to assess the quality of each virtual-queue realization (chromosome). The fitness value of the chromosome captures the cost of a potential schedule. The fitness value $f_{r,G}$ of a chromosome $r$ in generation $G$ is represented by the total waiting time of jobs that remain in the virtual queue.
\begin{equation}
  \label{equ:fitness}
  f_{r,G} = \sum_{i=1}^{l} (\omega_{i,j}^{\beta_j})
\end{equation}

% using equation~\ref{equ:wijbetaj}
The waiting time $\omega_{i,j}^{\beta_j}$ of the $i^{\text{th}}$ job waiting in the virtual queue of the $j^{\text{th}}$ tier should be calculated based on its order in the queue, as per the ordering $\beta_j$.
 %(default)width=\textwidth,height=0.82\textheight of whole page and changes when you change the figure % width=0.475\textwidth, height=0.475\textheight, width=0.65\textwidth % If we are to make it on the whole page..
%width=0.75\textwidth,height=0.62\textheight

The normalized fitness value of each schedule candidate is computed as follows:
\begin{equation}
  \label{equ:probabilityRouletteWheel}
  f_r  = \frac{f_{r,G}}{\sum_{C=1}^{n} (f_{C,G})}\;,\;\;\;r\!\in\!C
\end{equation}

Based on the normalized fitness values of the candidates, the Russian Roulette is used to select a set of schedule candidates to produce the next generation population, using the combination and mutation operators.

\subsection{Evolving the Scheduling Process}
\label{sec:step4}

To evolve a new population that holds new scheduling options for jobs in resource queues of the tier, the crossover and mutation genetic operators are both applied on randomly selected schedules (virtual queues) of the current generation. The crossover operator produces a new generation of virtual-queues from the current generation. The mutation operator applies random changes on a selected set of virtual-queues of the new generation to produce altered virtual-queues. These operators diversify the search direction into new search spaces to avoid getting stuck in a locally optimum solution. Overall, the \emph{Single-Point} crossover and \emph{Insert} mutation genetic operators are used in this paper. The rates of crossover and mutation operators are both set to be $0.1$ of the population size in each generation.

Figure~\ref{fig:GA} explains how each virtual-queue in a given generation are evolved to create a new virtual-queue of the next generation, using the crossover and mutation operators. Each chromosome (virtual-queue) represents a new scheduling of jobs. The jobs and their order of execution on the resource will be reflected by the segment of the virtual queue corresponding to the actual queue associated with the resource. As a result of the evolutionary process, each segment of the virtual queue corresponding to an actual queue will be in one of the following states:
\begin{itemize}
  \item Maintain the same set and the same ordering of jobs as in the previous generation.
  \item A new ordering for the same set of jobs as in the previous generation.
  \item A different set of jobs and their associated ordering.
\end{itemize}

For instance, queue $Q_{1,j}$ of \nolinebreak{\emph{Chromosome} ($n,\!1$)} in the first generation maintains exactly the same set and order of jobs in the second generation shown in queue $Q_{1,j}$ of \nolinebreak{\emph{Chromosome} ($n,\!2$)}. In contrast, queue $Q_{2,j}$ of \nolinebreak{\emph{Chromosome} ($1,\!1$)} in the first generation maintains the same set of jobs in the second generation, yet has got a new order of jobs as shown in queue $Q_{2,j}$ of \nolinebreak{\emph{Chromosome} ($1,\!2$)}. Finally, queue $Q_{2,j}$ of a random \nolinebreak{\emph{Chromosome} ($C,\!1$)} in the first generation has neither maintained the same set nor got the same order of jobs in the second generation shown in queue $Q_{2,j}$ of \nolinebreak{\emph{Chromosome} ($C,\!2$)}, which in turn would yield a new scheduling of jobs in the queue of resource $R_{2,j}$ if \nolinebreak{\emph{Chromosome} ($C,\!2$)} is later selected as the best chromosome of the tier-based genetic solution.

%% file: Results_Jan9.tex
\section{Experimental Work and Discussions on Results}
\label{sec:discResults}

The cloud environment adopted in this paper consists of two tiers, each of which has 3 computing resources. The jobs generated into the cloud environment are atomic and independent of each other. A job is first executed on one of the computing resources of the first tier and then moves for execution on one of the resources of the second tier. Each job is served by only one resource at a time, the scheduling strategy is non-preemptive.

Jobs arrive to the environment at the first tier and are queued in the arrival queue (tier dispatcher) of the environment. The arrival behaviour is modeled as a Poisson process. The running time of each job in a computing resource is assumed to be known in advance, generated with a rate $\nolinebreak{\mu\!=\!1}$ from the exponential distribution function $\nolinebreak{\text{exp}(\mu\!=\!1)}$~\cite{PerfEval_2016}. In each tier $T\!_j$, job migrations from a queue to another queue are permitted.

Two experiments are conducted. In the first experiment, we take advantage of the virtualized queue, and seek optimal schedules that produce minimum total waiting time among all jobs. Thus, the proposed genetic algorithm operates on all queues of the tier simultaneously. In the second experiment we apply the genetic algorithm on the individual queues of the tier. The penalty exponential scaling parameter $\nu$ is set to be $\nolinebreak{\nu\!=\!0.01}$, it is an arbitrary number used to visualize the penalty under different scheduling scenarios. In both experiments, we employ 10 chromosomes populations.

\subsection{Virtualized Queue Experiment}

The tier-based genetic solution is applied on the \emph{virtualized-queue}. The virtual-queue starts with an initial state that represents an initial scheduling $\beta_j$ of jobs in the tier (initial tier-state), which in turn yields an initial fitness and penalty of the virtual-queue. The initial fitness of the virtual-queue represents the total waiting time of jobs in the tier according to their initial scheduling in the virtual-queue. The tier-based genetic solution shown in Figure~\ref{fig:GA} is then applied on the virtual-queue (\emph{globally} at the tier level of the environment), which after some iterations finds a new enhanced scheduling of jobs in the virtual-queue (enhanced tier-state) that optimizes the objective function. The new enhanced tier-state yields a new improved fitness and penalty of the virtual-queue, which in turn is translated into a new enhanced scheduling of jobs in the resource queues of the tier that reduces the total waiting time and penalty of jobs \emph{globally} at the tier level of the environment.
\input{DRAWSummaryTierFigures_Jan9}

\input{Tier_based_Table_Jan9}

The results shown in Table~1 and Figure~\ref{fig:Case_TierBased} demonstrate the effectiveness of using the queue virtualization along with the tier-based genetic solution to reduce the total waiting time and thus the QoS violation penalty. The results of applying the tier-based genetic solution are reported in $6$ different instances. Figures~\ref{fig:TierBased_47_8462}-\ref{fig:TierBased_88_0743} are mapped to their corresponding first 3 instances of Table~1. For the virtual-queue of $19$ jobs shown in Table~1, the results show that the tier-based genetic solution has improved the fitness of the tier-state by $47.39\%$, reducing the total waiting time of jobs at each tier of the environment from $88.0743$ time units for the initial tier-state to $46.3381$ time units for the enhanced tier-state. The penalty amount has also been improved by $36.66\%$, reducing it from $0.586$ for the initial tier-state to $0.371$ for the enhanced tier-state.

Figure~\ref{fig:TierBased_88_0743} demonstrates the effectiveness of the tier-based genetic solution in reducing the total waiting time of jobs in the virtual-queue of $19$ jobs. The tier-based genetic solution has required $500$ iterations, each of which contains 10 chromosomes, to achieve the reported enhancement on the tier-state. A total of only $5000$ \emph{global} scheduling options for jobs in the tier is effectively explored in the search space of $19!$ (approximately $\nolinebreak{1.22\!\times\!10^{17}}$) different \emph{global} scheduling options at the tier level of the environment to improve the fitness and penalty of the tier-state by $47.39\%$ and $36.66\%$, respectively. Similarly, improvements are achieved with respect to the other 2 instances of the virtual-queue (12 and 15 jobs) shown in Table~1. Figures~\ref{fig:TierBased_47_8462} and \ref{fig:TierBased_50_8813}, respectively, depict such improvement.

In contrast, Figures~\ref{fig:TierBased_126_4679}-\ref{fig:TierBased_63_0545} are mapped to the second and third instances reported in Table~1. The tier-based genetic solution has required $1000$ iterations, each of which contains 10 chromosomes, to obtain the enhancement on the tier-state of each event. In this case, a virtual-queue of a large number of jobs has required more iterations so as to explore more \emph{global} scheduling options of the jobs at the tier level of the environment. For the virtual-queue of $31$ jobs shown in Table~1, the tier-based genetic solution has improved the performance of the tier by $25.64\%$ and $15.07\%$, respectively. Figure~\ref{fig:TierBased_126_4679} shows that a total of only $10,\!000$ out of $31!$ (approximately $\nolinebreak{8.22\!\times\!10^{33}}$) possible \emph{global} scheduling options for jobs at the tier level of the environment are effectively explored to achieve the latter enhancements. Similarly, performance improvements are achieved with respect to the other 2 instances of the virtual-queue (32 and 27 jobs) shown in Table~1, and their corresponding performance are depicted in Figures~\ref{fig:TierBased_217_1755} and \ref{fig:TierBased_63_0545}, respectively.

\subsection{Segmented Queue Experiment}

The genetic solution is applied at \emph{each individual queue} level. Each one of the three queues holds an initial set of jobs to be executed on the resource associated with that queue. The waiting time of each job is calculated based on its position in the queue. The proposed genetic algorithm is then used to seek an optimal ordering of the jobs that are queued for execution by the resource associated with that queue, such that the total waiting time of these jobs is minimized. The genetic algorithm in this case seeks an optimal schedule in a reduced search space, since the optimal ordering is sought on each queue individually. In other words, a genetic search strategy is performed on each queue. The total waiting time, of all jobs in the three queues, are computed.

\input{DRAWSummaryQueueFigures_Jan9}

Table~2 shows the results of applying the genetic algorithm on the three resource queues, in two different instances. The first instance represents a job allocation whereby resource-1 is allocated 14 jobs, resource-2 16 jobs, and resource-3 15 jobs. The second instance represents a job allocation whereby resource-1 is allocated 19 jobs, resource-2 23 jobs, and resource-3 14 jobs. Table~2 enumerates the total number of \emph{local} orderings (schedules) for the first instance. There are $14!$ possible orderings for queue-1, $16!$ for queue-2, and $15!$ for queue-3.

\input{Queue_based_Table_Jan9}

The table shows a $36.04\%$ improvement from the initial ordering for queue-1, a reduction from $154.1339$ time units of total waiting time to $98.5818$ time units of total waiting time. The QoS violation penalty has improved by $20.24\%$, from $0.786$ due to the initial ordering, to $0.627$ due to the improved ordering computed by the genetic search strategy.

Figure~\ref{fig:Server1_154.1339} depicts the total waiting time of jobs allocated to resource-1 during the search process. After $150$ genetic iterations, an optimal solution was found. Each iteration 10 chromosomes are used to evolve the optimal schedule. Thus, $1500$ orderings are constructed and genetically manipulated throughout the search process, as apposed to $14!$, if we were to employ a brute-force search strategy.  Similar observations are in order with respect to resource-2 and resource-3, as can be seen in the figure.

Table~2 reveals the magnitude of search space growth as a result of increasing the number of jobs allocated a given resource. For example, if we consider the impact of increasing the number of jobs allocated to resource-1 from $14$ jobs to $23$ jobs. In a brute-force search strategy, the search space will increase from $14!$ to $23!$. In contrast, the genetic search strategy needed to expand the search space from $1500$ populations to $7500$ populations. After $7500$ genetic iterations the waiting time was improved by $58.16\%$ from the initial ordering. The total waiting time of jobs was reduced from $208.596$ waiting time units due to the initial job ordering to $87.2667$ waiting time units due to the genetically improved ordering. Figures~\ref{fig:Server1_150.8208}-\ref{fig:Server3_145.0253} demonstrate the effectiveness of using the queue-based genetic solution to decrease the total waiting time of jobs in the three resources: resource-1, resource-2, and resource-3, respectively.

\subsection{Comparison}

\begin{figure}[!ht]
\centering \includegraphics[width=0.5\textwidth,height=0.2\textheight]{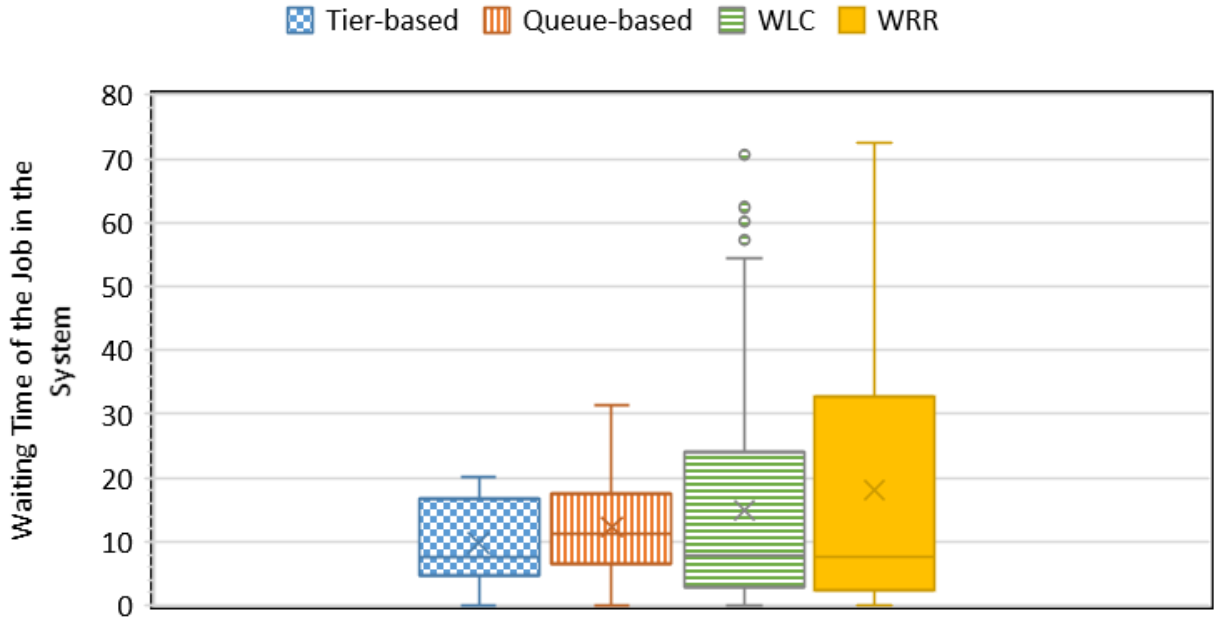}
	 \caption{Maximum Waiting Time Performance Comparison}
     \label{fig:Paper1_Comparison}
\end{figure}

Figure~\ref{fig:Paper1_Comparison} and Table~3 contrast the performance of both genetic strategies, i.e, the virtualized queue search strategy and the individualized queue strategy. The initial orderings of the three queues, and by implication, that of the virtualized queue are the same. WRR's based ordering entailed $3617$ units of total waiting time. WLC's based ordering entailed $3001$ units of total waiting time. The individualized queue genetic search strategy was produced an ordering that entails $2464$ units of waiting time, a $32\%$ reduction compared to the WRR strategy and $18\%$ reduction compared to the WLC strategy. The virtualized queue genetic search strategy produced an ordering that entails $1961$ units of waiting time. That is a reduction of $46\%$ compared to he WRR strategy and $35\%$ reduction compared to the WLC strategy.
\begin{table}[!t]
\centering
\label{tab:tableComparison}
\caption{Total Waiting Time of Jobs in each Approach}
\begin{tabular}{clclcc}
\hline
\textbf{Virtualized Queue} & \textbf{Segmented Queue} & \textbf{WLC} & \textbf{WRR} \\ \hline
1961.34                    & 2464.61                  & 3001.82      &  3617.95     \\ \hline
\end{tabular}
\end{table}
 %(default)width=0.5\textwidth,height=0.2\textheight of whole page and changes when you change the figure
%width=0.475\textwidth,height=0.19\textheight

Figure~\ref{fig:Paper1_Comparison} depicts the average waiting performance of the four scheduling strategies. The virtualized queue genetic strategy has produced the shortest average waiting time per job, with an average waiting time of $10$ time units. The individualized queue search strategy produced an average waiting time of $13$ time units. The WRR and WLC job ordering strategies delivered inferior performance.

On the other hand, the individualized queue strategy has yielded a maximum job waiting time of $19$ time units. The WRR produced a maximum job waiting time of $32$ time units, while in the WLC produced a maximum job waiting time of $24$. The virtualized queue scheduling strategy delivered a maximum job waiting time of $16$ time units. Overall, the virtualized queue scheduling strategy delivered the best performance in minimizing the total waiting time and thus the lowest QoS penalty.

%% file: DRAWSummaryTierFigures_Jan9.tex
\begin{figure*}[!ht]
        \centering
        \begin{subfigure}{0.321\textwidth}
		        \includegraphics[width=\textwidth]
                {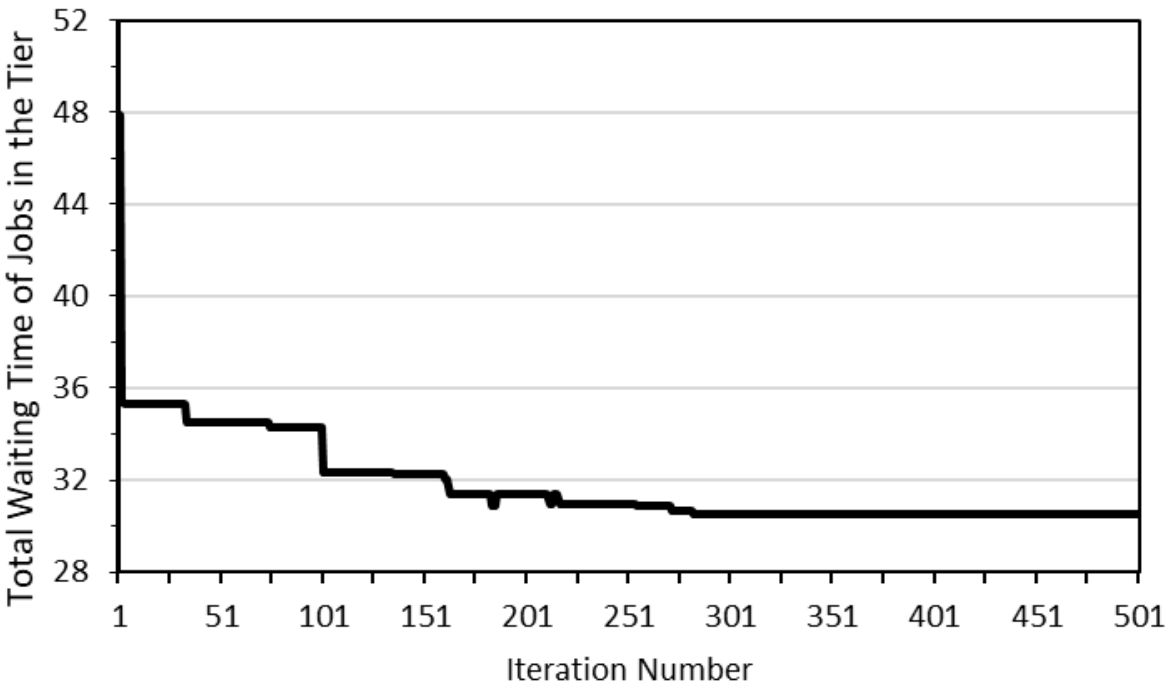}
	            \caption{Virtual-Queue of 12 Jobs}
	            \label{fig:TierBased_47_8462} %,height=0.205\textheight
        \end{subfigure}
        ~
        \begin{subfigure}{0.321\textwidth}
		        \includegraphics[width=\textwidth]
                {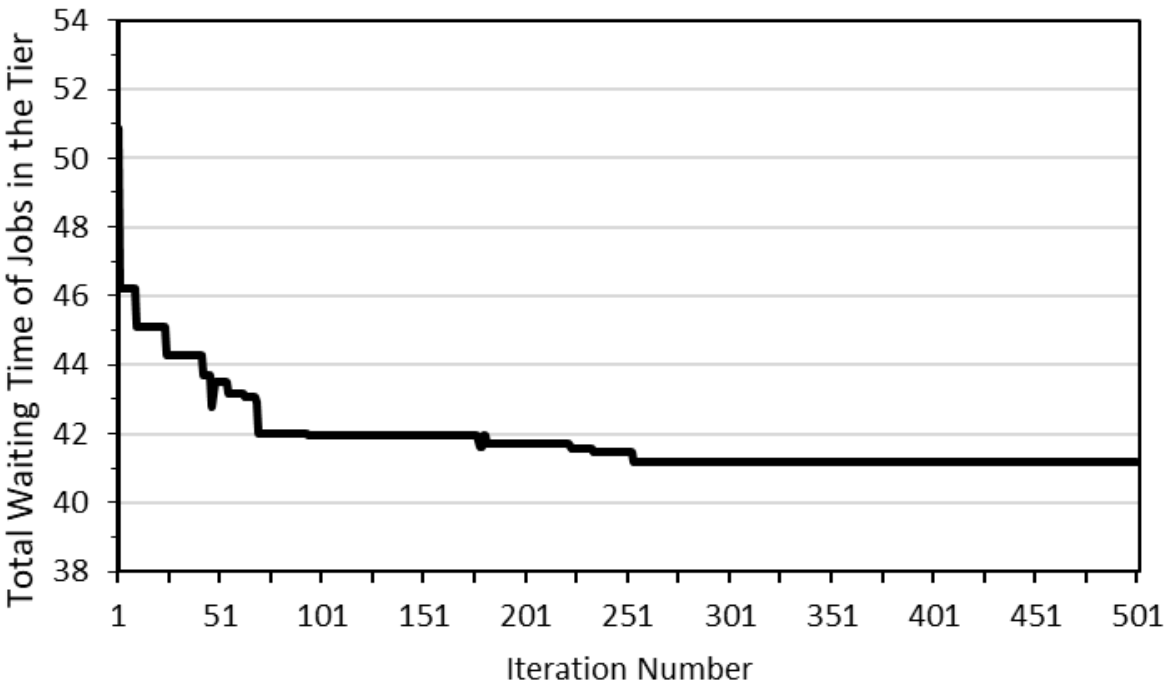}
	            \caption{Virtual-Queue of 15 Jobs}
	            \label{fig:TierBased_50_8813} %,height=0.205\textheight
        \end{subfigure}
        ~
        \begin{subfigure}{0.321\textwidth}
                \includegraphics[width=\textwidth]
                {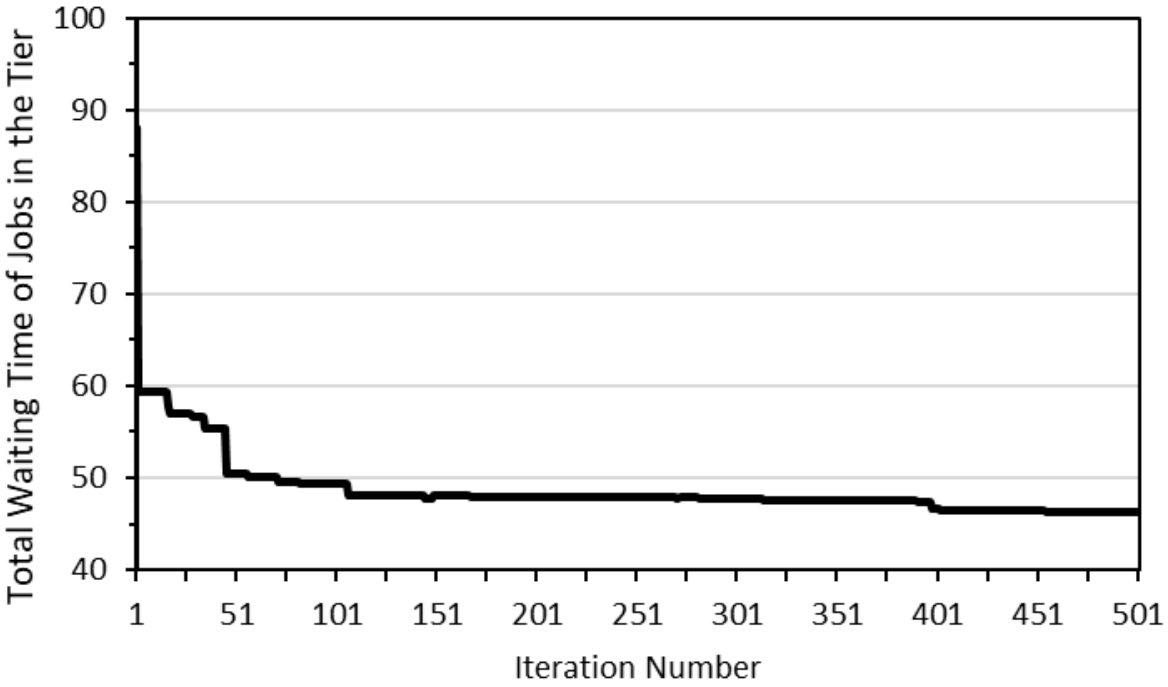}
	            \caption{Virtual-Queue of 19 Jobs}
	            \label{fig:TierBased_88_0743}
        \end{subfigure}

        \begin{subfigure}{0.321\textwidth}
		        \includegraphics[width=\textwidth]
                {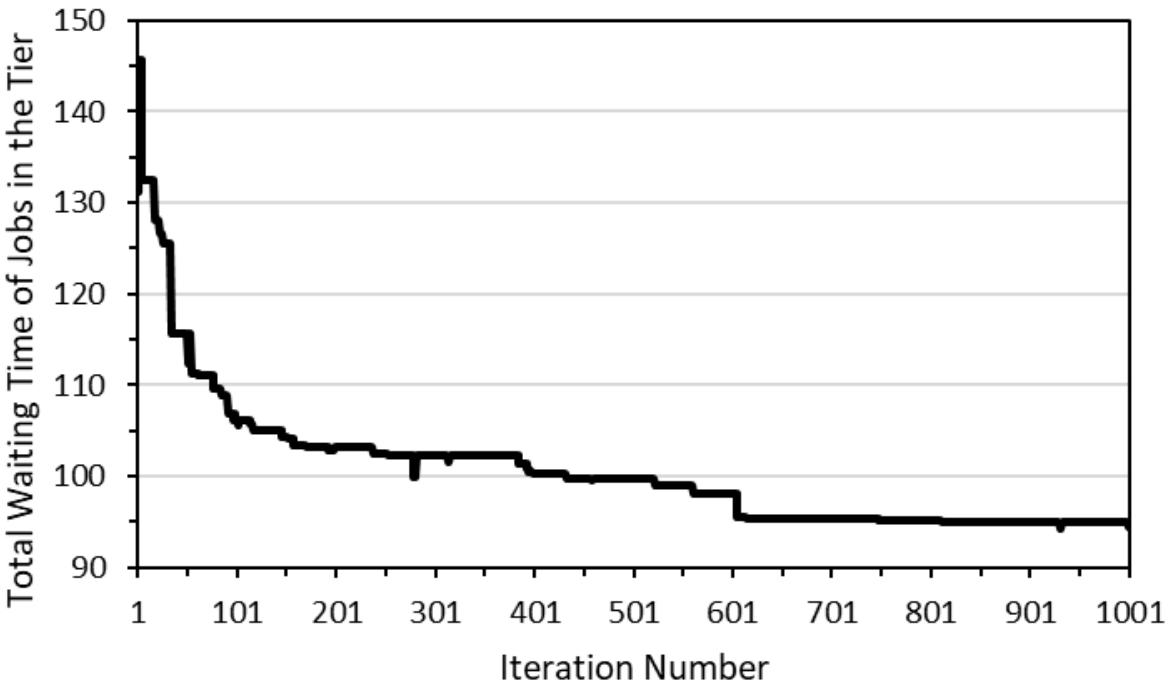}
	            \caption{Virtual-Queue of 31 Jobs}
	            \label{fig:TierBased_126_4679} %,height=0.205\textheight
        \end{subfigure}
        ~
        \begin{subfigure}{0.321\textwidth}
		        \includegraphics[width=\textwidth]
                {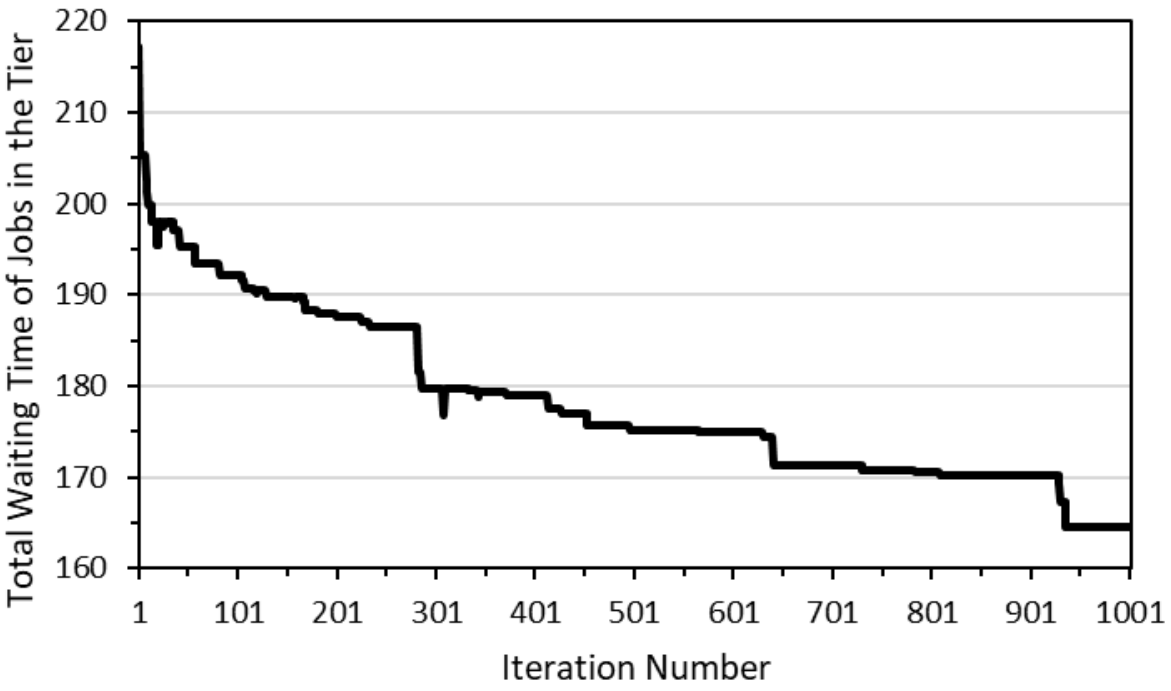}
	            \caption{Virtual-Queue of 32 Jobs}
	            \label{fig:TierBased_217_1755} %,height=0.205\textheight
        \end{subfigure}
        ~
        \begin{subfigure}{0.321\textwidth}
                \includegraphics[width=\textwidth]
                {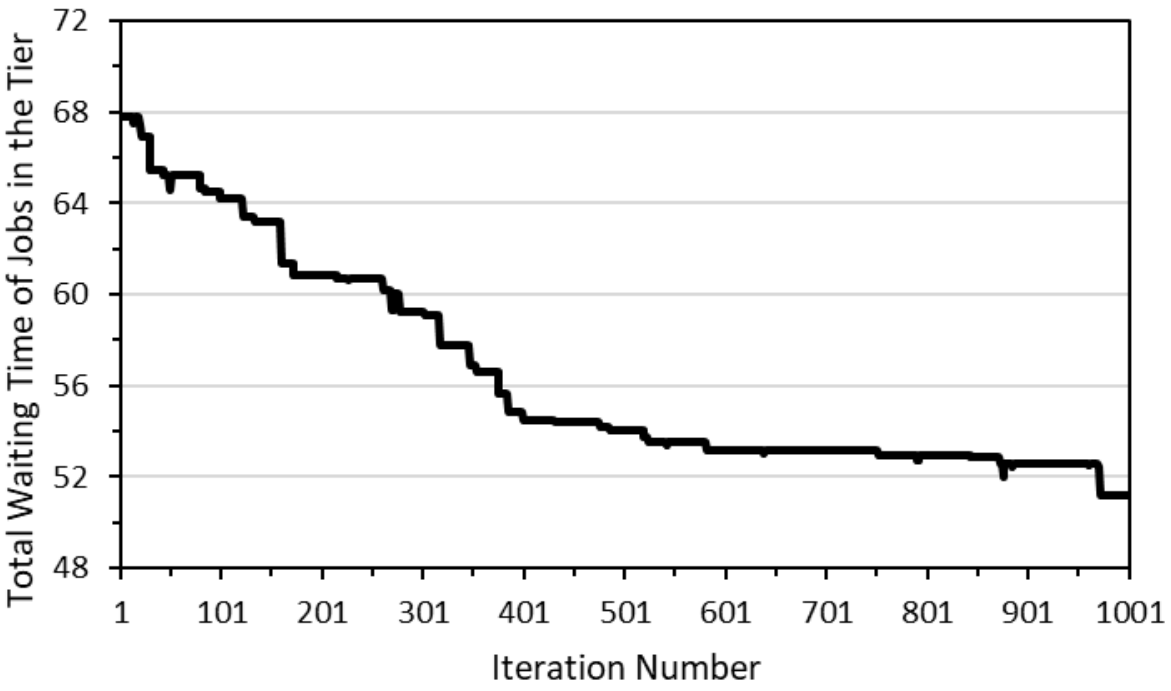}
	            \caption{Virtual-Queue of 27 Jobs}
	            \label{fig:TierBased_63_0545}
        \end{subfigure}

        \caption{Tier-based Scheduling}
        \label{fig:Case_TierBased}
\end{figure*} % \begin{subfigure}{0.47\textwidth} [width=\textwidth,height=0.205\textheight]

%% file: Tier_based_Table_Jan9.tex
%centering

\begin{table*}[!ht]
\label{tab:TierBasedTable}
\caption{Tier-based Scheduling}
\begin{center}\scalebox{0.915}{
\begin{threeparttable}
\begin{tabular}{c|c|cccc|cc}
\hline
\multirow{2}{*}{}
& \multirow{2}{*}{\textbf{\begin{tabular}[c]{@{}c@{}}Virtual-Queue \\ Length\end{tabular}}}\tnote{1}
& \multicolumn{2}{c}{\textbf{Initial}\tnote{2}} & \multicolumn{2}{c|}{\textbf{Enhanced}\tnote{3}}
& \multicolumn{2}{c}{\textbf{Improvement}}  \\ \cline{3-8}
& & \textbf{Waiting}  & \textbf{Penalty} & \textbf{Waiting} & \textbf{Penalty} & \textbf{Waiting \%} & \textbf{Penalty \%} \\ \hline \hline

Figure\ref{fig:TierBased_47_8462}      & 12   & 47.8462   & 0.380   & 30.4821   & 0.263   & 36.29\%   & 30.90\%       \\
Figure\ref{fig:TierBased_50_8813}      & 15   & 50.8813   & 0.399   & 41.1748   & 0.338   & 19.08\%   & 15.37\%       \\
Figure\ref{fig:TierBased_88_0743}      & 19   & 88.0743   & 0.586   & 46.3381   & 0.371   & 47.39\%   & 36.66\%       \\ \hline

Figure\ref{fig:TierBased_126_4679}     & 31   & 126.4679  & 0.718   & 94.0426   & 0.610   & 25.64\%   & 15.07\%       \\
Figure\ref{fig:TierBased_217_1755}     & 32   & 217.1755  & 0.886   & 164.4844  & 0.807   & 24.26\%   & 8.92\%        \\
Figure\ref{fig:TierBased_63_0545}      & 27   & 63.0545   & 0.468   & 51.2031   & 0.401   & 18.80\%   & 14.32\%       \\ \hline

\end{tabular}
\begin{tablenotes}\footnotesize%\scriptsize
\item[1] \textbf{Virtual-Queue Length} represents the total number of jobs in queues of the tier. For instance, the first entry of the table (12) means that the 3 queues of the tier all together contain 12 jobs.
\item[2] \textbf{Initial Waiting} represents the total waiting time of jobs in the virtual-queue according to the initial scheduling of jobs before using the tier-based genetic solution.
\item[3] \textbf{Enhanced Waiting} represents the total waiting time of jobs in the virtual-queue according to the final/enhanced scheduling of jobs found after using the tier-based genetic solution.
\end{tablenotes}
\end{threeparttable}}
\end{center}
\end{table*}

%% file: DRAWSummaryQueueFigures_Jan9.tex
\begin{figure*}[!ht]
        \centering
        \begin{subfigure}{0.32\textwidth}
		        \includegraphics[width=\textwidth]
                {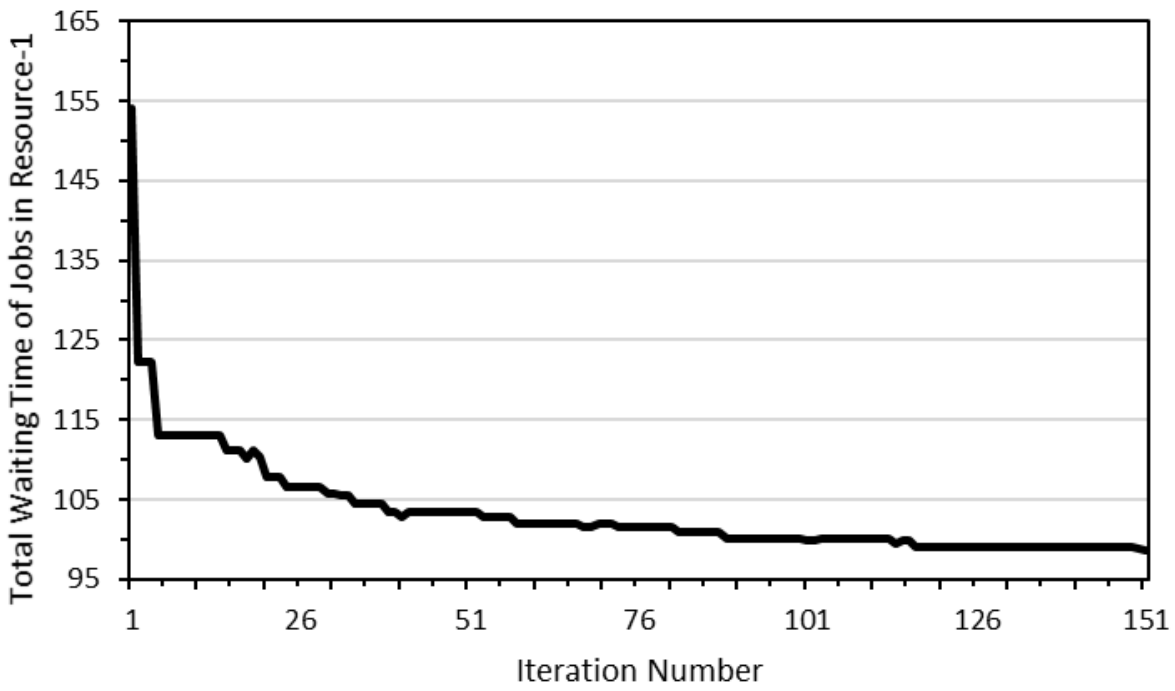}
	            \caption{Resource 1 (Queue of 14 Jobs)}
	            \label{fig:Server1_154.1339} %\includegraphics[width=\textwidth,height=0.46\textheight]
        \end{subfigure}
        ~
        \begin{subfigure}{0.32\textwidth}
		        \includegraphics[width=\textwidth]
                {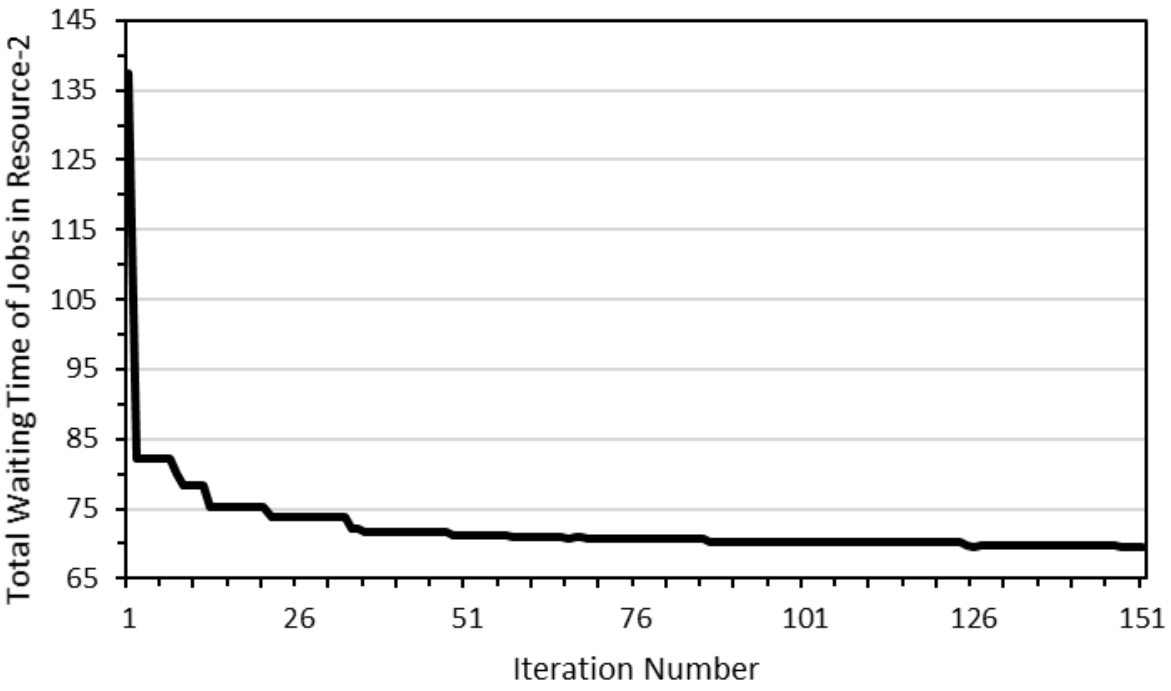}
	            \caption{Resource 2 (Queue of 16 Jobs)}
	            \label{fig:Server2_137.3684} %,height=0.205\textheight
        \end{subfigure}
        ~
        \begin{subfigure}{0.32\textwidth}
                \includegraphics[width=\textwidth]
                {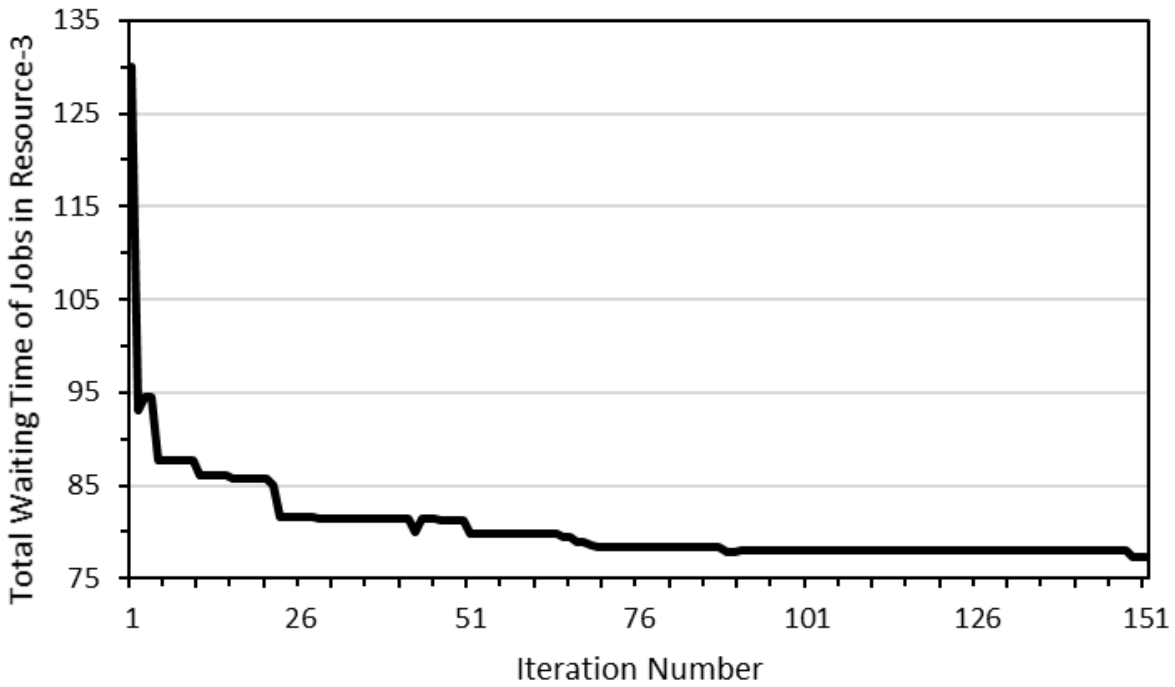}
	            \caption{Resource 3 (Queue of 15 Jobs)}
	            \label{fig:Server3_130.0566}
        \end{subfigure}

%%%%%%%%%%%%%%%%%%%%%%%%%%%%%%%%%%%%%%%%%%%%%%%%%%

        \begin{subfigure}{0.32\textwidth}
		        \includegraphics[width=\textwidth]
                {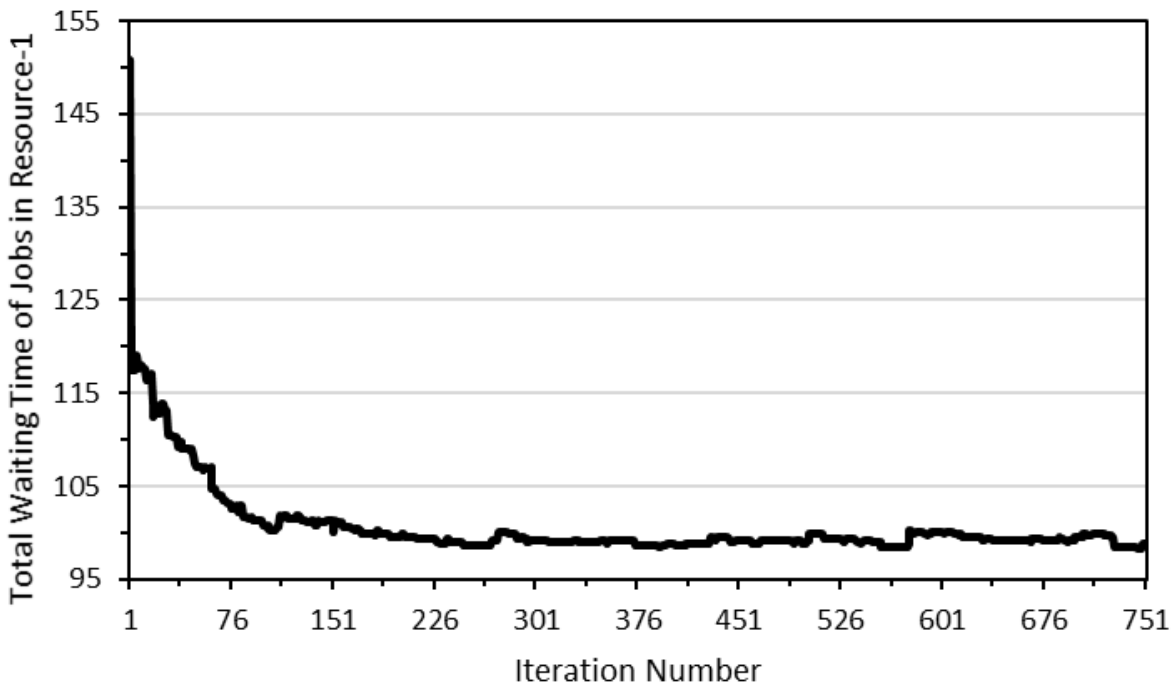}
	            \caption{Resource 1 (Queue of 19 Jobs)}
	            \label{fig:Server1_150.8208} %,height=0.205\textheight
        \end{subfigure}
        ~
        \begin{subfigure}{0.32\textwidth}
		        \includegraphics[width=\textwidth]
                {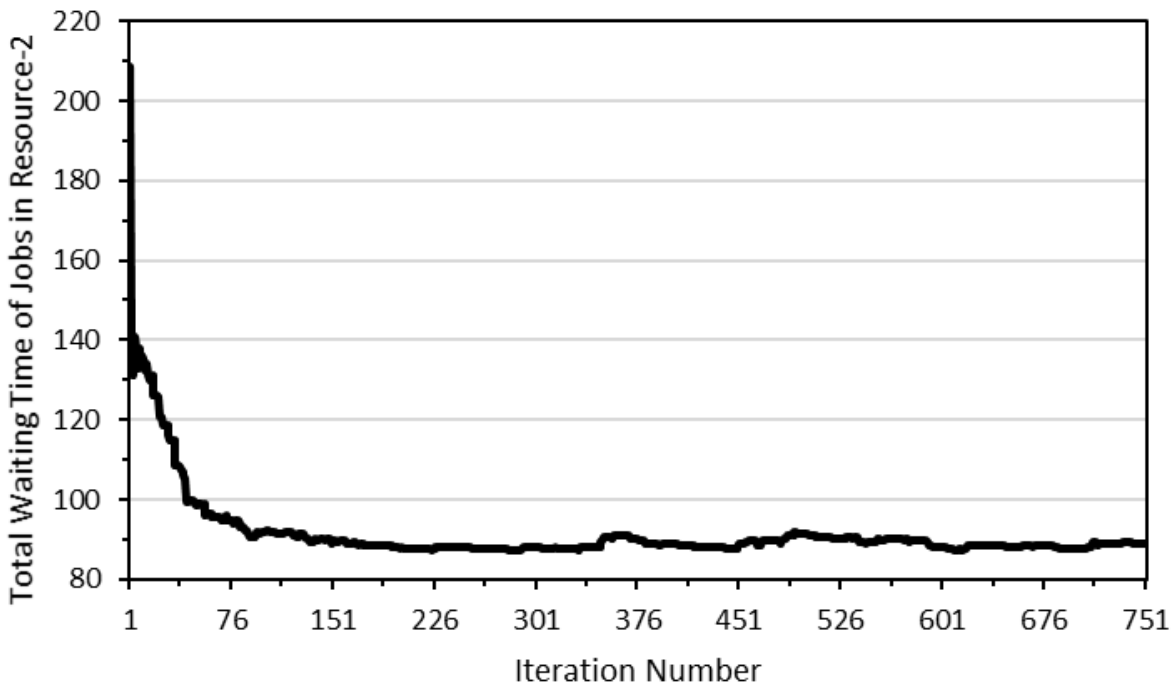}
	            \caption{Resource 2 (Queue of 23 Jobs)}
	            \label{fig:Server2_208.5960} %,height=0.205\textheight
        \end{subfigure}
        ~
        \begin{subfigure}{0.32\textwidth}
                \includegraphics[width=\textwidth]
                {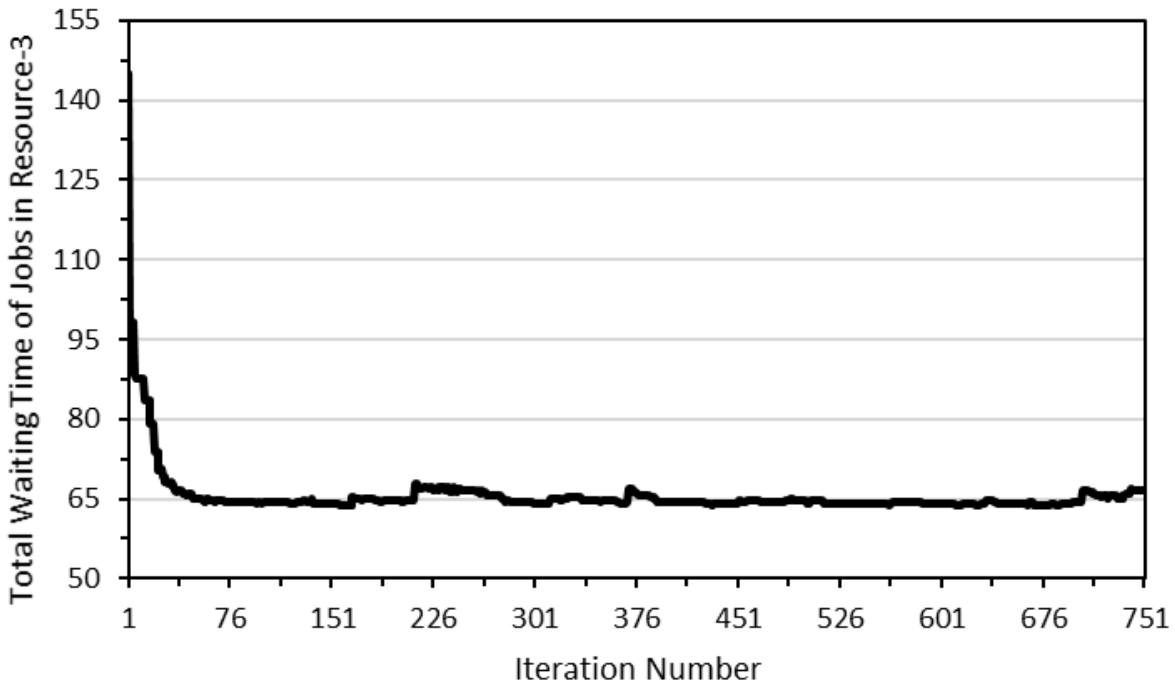}
	            \caption{Resource 3 (Queue of 14 Jobs)}
	            \label{fig:Server3_145.0253}
        \end{subfigure}

        \caption{Queue-based Scheduling}
        \label{fig:Case_QueueBased}
\end{figure*} % \begin{subfigure}{0.47\textwidth} [width=\textwidth,height=0.205\textheight]

%% file: Queue_based_Table_Jan9.tex
\begin{table*}[!ht]
\label{tab:QueueTable}
\caption{Queue-based Scheduling}
\begin{center}\scalebox{0.88}{
\begin{threeparttable}
\begin{tabular}{c|c|cccc|cc}
\hline
\multirow{2}{*}{}
& \multirow{2}{*}{\textbf{\begin{tabular}[c]{@{}c@{}}Queue \\ Length\end{tabular}}}\tnote{4}
& \multicolumn{2}{c}{\textbf{Initial}\tnote{5}} & \multicolumn{2}{c|}{\textbf{Enhanced}\tnote{6}}
& \multicolumn{2}{c}{\textbf{Improvement}}  \\ \cline{3-8}
& & \textbf{Waiting}  & \textbf{Penalty} & \textbf{Waiting} & \textbf{Penalty} & \textbf{Waiting \%} & \textbf{Penalty \%} \\ \hline \hline

Resource 1 Figure\ref{fig:Server1_154.1339}  & 14 & 154.1339 & 0.786 & 98.5818  & 0.627 & 36.04\% & 20.24\%   \\
Resource 2 Figure\ref{fig:Server2_137.3684}  & 16 & 137.3684 & 0.747 & 69.4641  & 0.501 & 49.43\% & 32.95\%   \\
Resource 3 Figure\ref{fig:Server3_130.0566}  & 15 & 130.0566 & 0.728 & 77.3358  & 0.539 & 40.54\% & 25.99\%   \\ \hline\hline

Resource 1 Figure\ref{fig:Server1_150.8208}  & 19 & 150.8208 & 0.779 & 98.1834  & 0.625 & 34.90\% & 19.69\%   \\
Resource 2 Figure\ref{fig:Server2_208.5960}  & 23 & 208.596  & 0.876 & 87.2667  & 0.582 & 58.16\% & 33.53\%   \\
Resource 3 Figure\ref{fig:Server3_145.0253}  & 14 & 145.0253 & 0.765 & 63.8502  & 0.472 & 55.97\% & 38.35\%   \\ \hline

\end{tabular}
\begin{tablenotes}\footnotesize%\scriptsize
\item[4] \textbf{Queue Length} represents the number of jobs in the queue of a resource.
\item[5] \textbf{Initial Waiting} represents the total waiting time of jobs in the queue according to the initial scheduling of jobs before using the queue-based genetic solution.
\item[6] \textbf{Enhanced Waiting} represents the total waiting time of jobs in the queue according to the final/enhnaced scheduling of jobs found after using the queue-based genetic solution.
\end{tablenotes}
\end{threeparttable}}
\end{center}
\end{table*}

%% file: Conclusion_Jan9.tex
\section{Conclusion}
\label{sec:conc}

This paper presents a genetic algorithm for tackling the job scheduling problem in a multi-tier cloud computing environment. The paper makes the connection between penalties payable due to QoS violations and job waiting time. This connection establishes a framework for facilitating penalty management and mitigation that service providers can utilize in high demand/limited resources situations. It is assumed that each tier of the environment consists of a set of identical computing resources. A queue is associated with each one of these resources.
To achieve maximum resource utilization and minimum waiting time, a virtualized queue abstraction is proposed. Each virtual queue realization represents an execution ordering of jobs. This virtualized queue abstraction collapses the search spaces of all queues into one search space of orderings, and thus allowing the genetic algorithm to seek optimal schedules at the tier level.
The paper presented experimental work to investigate the performance of the proposed biologically inspired strategy to WRR and WLC, as well as an individualized queue strategy. It is concluded that the proposed job scheduling strategy delivers performance that is superior to that of both WRR and WLC. The genetic search strategy when applied at the individual queue delivers performance also superior to that of WRR and WLC. However, the genetic search strategy applied at the virtual queue still delivered the best performance compared to all the other search strategies.

\section{Future Work}
\label{sec:futureW}

The proposed scheduling strategy does not contemplate the impact of schedules optimized in a given tier on the performance of schedules on the subsequent tiers. Therefore, it is the intent of the authors to expand the work reported in this paper to investigate such impact and to extend the algorithms proposed in this paper so as to mitigate the impact of tier dependency. Furthermore, the formulation presented in this paper treats the penalty factor of each job as a function of time to be identical. Typically, cloud computing jobs tend to vary with respect to the QoS violation penalties. Therefore, it is imperative to modify the penalty model so as to reflect such sensitivity so as to force the scheduling process to produce minimum penalty schedules, and not necessarily minimum total waiting time schedules.

%% file: Paper1_Jan9.bbl
\begin{thebibliography}{10}
\providecommand{\url}[1]{#1}
\csname url@samestyle\endcsname
\providecommand{\newblock}{\relax}
\providecommand{\bibinfo}[2]{#2}
\providecommand{\BIBentrySTDinterwordspacing}{\spaceskip=0pt\relax}
\providecommand{\BIBentryALTinterwordstretchfactor}{4}
\providecommand{\BIBentryALTinterwordspacing}{\spaceskip=\fontdimen2\font plus
\BIBentryALTinterwordstretchfactor\fontdimen3\font minus
  \fontdimen4\font\relax}
\providecommand{\BIBforeignlanguage}[2]{{%
\expandafter\ifx\csname l@#1\endcsname\relax
\typeout{** WARNING: IEEEtran.bst: No hyphenation pattern has been}%
\typeout{** loaded for the language `#1'. Using the pattern for}%
\typeout{** the default language instead.}%
\else
\language=\csname l@#1\endcsname
\fi
#2}}
\providecommand{\BIBdecl}{\relax}
\BIBdecl

\bibitem{1_cloudRaouf_2010}
Q.~Zhang, L.~Cheng, and R.~Boutaba, ``Cloud computing: state-of-the-art and
  research challenges,'' \emph{Journal of Internet Services and Applications},
  vol.~1, no.~1, pp. 7--18, 2010.

\bibitem{cloudSurv2012}
Y.~Jadeja and K.~Modi, ``Cloud computing - concepts, architecture and
  challenges,'' in \emph{Proceedings of the International Conference on
  Computing, Electronics and Electrical Technologies}, March 2012, pp.
  877--880.

\bibitem{JobSizeSched_2000}
K.~Aida, ``Effect of job size characteristics on job scheduling performance,''
  in \emph{Proceedings of the Workshop on Job Scheduling Strategies for
  Parallel Processing}, May 2000, pp. 1--17.

\bibitem{SoftHardSLA}
H.~J. Moon, Y.~Chi, and H.~Hacigumus, ``Performance evaluation of scheduling
  algorithms for database services with soft and hard {SLAs},'' in
  \emph{Proceedings of the Second International Workshop on Data Intensive
  Computing in the Clouds}, November 2011, pp. 81--90.

\bibitem{SLAlevels}
G.~Stavrinides and H.~Karatza, ``The effect of workload computational demand
  variability on the performance of a {SaaS} cloud with a multi-tier {SLA},''
  in \emph{Proceedings of the Fivth {IEEE} International Conference on Future
  Internet of Things and Cloud}, August 2017, pp. 10--17.

\bibitem{A_Comparative_Survey_2014}
H.~Shoja, H.~Nahid, and R.~Azizi, ``A comparative survey on load balancing
  algorithms in cloud computing,'' in \emph{Proceedings of the International
  Conference on Computing, Communication and Networking Technologies}, July
  2014, pp. 1--5.

\bibitem{LBghomiSurv2017}
E.~Ghomi, A.~Rahmani, and N.~Qader, ``Load-balancing algorithms in cloud
  computing: A survey,'' \emph{Journal of Network and Computer Applications},
  vol.~88, pp. 50--71, 2017.

\bibitem{TaskSurv_2019}
A.~Arunarani, D.~Manjula, and V.~Sugumaran, ``Task scheduling techniques in
  cloud computing: A literature survey,'' \emph{Future Generation Computer
  Systems}, vol.~91, pp. 407--415, 2019.

\bibitem{Bianca_2004}
B.~Schroeder and M.~Harchol-Balter, ``Evaluation of task assignment policies
  for supercomputing servers: The case for load unbalancing and fairness,''
  \emph{Journal of Cluster Computing}, vol.~7, no.~2, pp. 151--161, 2004.

\bibitem{MinMin_2013}
G.~Liu, J.~Li, and J.~Xu, ``An improved {Min-Min} algorithm in cloud
  computing,'' in \emph{Proceedings of the International Conference of Modern
  Computer Science and Applications}, May 2013, pp. 47--52.

\bibitem{MaxMin_2014}
X.~Li, Y.~Mao, X.~Xiao, and Y.~Zhuang, ``An improved {Max-Min} task-scheduling
  algorithm for elastic cloud,'' in \emph{Proceedings of the International
  Symposium on Computer, Consumer and Control}, June 2014, pp. 340--343.

\bibitem{Mor_2011}
V.~Gupta, M.~Harchol-Balter, K.~Sigman, and W.~Whitt, ``Analysis of
  join-the-shortest-queue routing for web server farms,'' \emph{Performance
  Evaluation}, vol.~64, no. 9--12, pp. 1062--1081, 2007.

\bibitem{WLC_2015}
L.~Kang and X.~Ting, ``Application of adaptive load balancing algorithm based
  on minimum traffic in cloud computing architecture,'' in \emph{Proceedings of
  the International Conference on Logistics, Informatics and Service Sciences},
  July 2015, pp. 1--5.

\bibitem{WRR_2014}
W.~Wang and G.~Casale, ``Evaluating weighted round robin load balancing for
  cloud web services,'' in \emph{Proceedings of the International Symposium on
  Symbolic and Numeric Algorithms for Scientific Computing}, September 2014,
  pp. 393--400.

\bibitem{JIQ_2011}
Y.~Lu, Q.~Xie, G.~Kliot, A.~Geller, J.~Larus, and A.~Greenberg,
  ``{Join-Idle-Queue}: A novel load balancing algorithm for dynamically
  scalable web services,'' \emph{Performance Evaluation}, vol.~68, no.~11, pp.
  1056--1071, 2011.

\bibitem{Pooja}
P.~Samal and P.~Mishra, ``Analysis of variants in round robin algorithms for
  load balancing in cloud computing,'' \emph{International Journal of Computer
  Science and Information Technologies}, vol.~4, no.~3, pp. 416--419, 2013.

\bibitem{heuristic1}
K.~Boloor, R.~Chirkova, T.~Salo, and Y.~Viniotis, ``Heuristic-based request
  scheduling subject to a percentile response time {SLA} in a distributed
  cloud,'' in \emph{Proceedings of the {IEEE} Global Telecommunications
  Conference}, December 2010, pp. 1--6.

\bibitem{ACO_MaxMin_2015}
N.~Ghumman and R.~Kaur, ``Dynamic combination of improved max-min and ant
  colony algorithm for load balancing in cloud system,'' in \emph{Proceedings
  of the International Conference on Computing, Communication and Networking
  Technologies}, July 2015, pp. 1--5.

\bibitem{PSO_2014}
F.~Ramezani, J.~Lu, and F.~Hussain, ``Task-based system load balancing in cloud
  computing using particle swarm optimization,'' \emph{International Journal of
  Parallel Programming}, vol.~42, no.~5, pp. 739--754, 2014.

\bibitem{heuristicPSO}
M.~Nouiri, A.~Bekrar, A.~Jemai, S.~Niar, and A.~Ammari, ``An effective and
  distributed particle swarm optimization algorithm for flexible job-shop
  scheduling problem,'' \emph{Journal of Intelligent Manufacturing}, vol.~29,
  no.~3, pp. 603--615, 2018.

\bibitem{Mateos2013AnAA}
C.~Mateos, E.~Pacini, and C.~Garino, ``An {ACO}-inspired algorithm for
  minimizing weighted flowtime in cloud-based parameter sweep experiments,''
  \emph{Advances in Engineering Software}, vol.~56, pp. 38--50, 2013.

\bibitem{Pandey2010APS}
S.~Pandey, L.~Wu, S.~Guru, and R.~Buyya, ``A particle swarm optimization-based
  heuristic for scheduling workflow applications in cloud computing
  environments,'' \emph{Proceedings of the IEEE International Conference on
  Advanced Information Networking and Applications}, pp. 400--407, 2010.

\bibitem{GA3}
Y.~Xiaomei, Z.~Jianchao, L.~Jiye, and L.~Jiahua, ``A genetic algorithm for job
  shop scheduling problem using co-evolution and competition mechanism,'' in
  \emph{Proceedings of the International Conference on Artificial Intelligence
  and Computational Intelligence}, October 2010, pp. 133--136.

\bibitem{GaTabu1}
X.~Li and L.~Gao, ``An effective hybrid genetic algorithm and tabu search for
  flexible job shop scheduling problem,'' \emph{International Journal of
  Production Economics}, vol. 174, no.~4, pp. 93--110, 2016.

\bibitem{PerfEval_2016}
T.~Atmaca, T.~Begin, A.~Brandwajn, and H.~Castel-Taleb, ``Performance
  evaluation of cloud computing centers with general arrivals and service,''
  \emph{IEEE Transactions on Parallel and Distributed Systems}, vol.~27, no.~8,
  pp. 2341--2348, 2016.

\end{thebibliography}
